\documentclass[pra,amsmath,amssymb]{revtex4}
\usepackage{braket}
\usepackage{graphicx}
\usepackage{subfigure}
\def\beq{\begin{equation}}
\def\eeq{\end{equation}}
\def \mA{\mathbb{A}}
\def \mB{\mathbb{B}}

\def \mU{\mathbb{U}}

\def \mK{\mathbb{K}}
\def \mN{\mathbb{N}}
\def \mI{\mathbb{I}}
\def \mF{\mathbb{F}}
\def \gf{\tilde{g}}

\def \mP{\mathbb{P}}
\def \mH {\mathbb{H}}
\begin{document}
\title{Entanglement Transfer from Bosonic Systems to Qubits }
\author{Andrzej Veitia}
\email{aveitia@physics.miami.edu}
\affiliation{Department of Physics, University of Miami, Coral Gables, 33146, FL, USA}

\begin{abstract}
We study the entanglement of a pair of qubits resulting from their interaction with a bosonic system. Here we restrict our discussion to the case where the set of operators acting on different qubits commute. A special class of interactions inducing entanglement in  an initially separable two qubit system is discussed.
Our results apply to the case where the initial state of the  bosonic system is represented by a statistical mixture of states with fixed particle number.
\end{abstract}\maketitle
\section{\label{Intro} Introduction}

\begin{figure}[htb]
\centering
\subfigure[Entanglement Transfer from 2-qubit system $B_{1}B_{2}$ to 2-qubit system $A_{1}A_{2}$]
{
\includegraphics[scale=0.8]{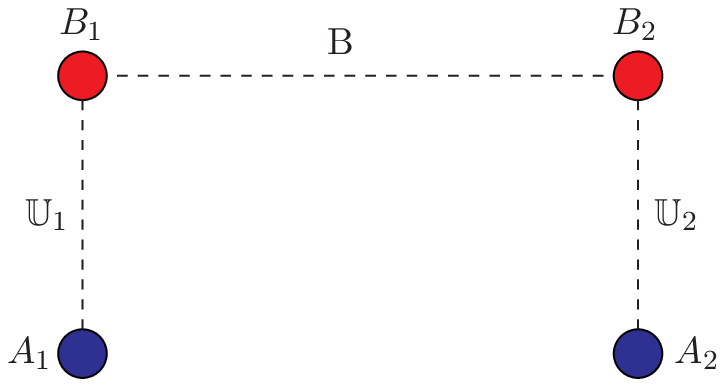}
\label{F:qubit-qubit}
}
\hspace{2.0 cm}
\subfigure[Generalized entanglement transfer. Here the operators coupling $A_{1}$ to B and $A_{2}$ to B commute]{
\includegraphics[scale=0.8]{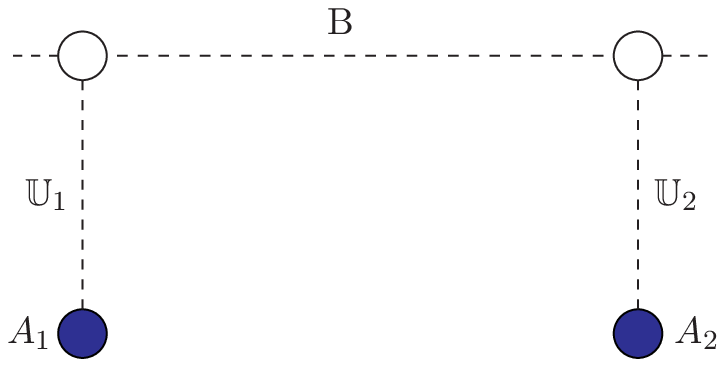}
\label{F:qubit-background}
}
\caption[]{ Entanglement transfer schemes.}
\end{figure}

The quantum correlations between subsystems present in entangled states are indispensable for many quantum communication protocols \cite{Chuang}. However, these correlations
cannot be created by local operations and classical communication(LOCC). Therefore, in order to entangle two systems $A_{1}$ and $A_{2}$, it is necessary to apply a global operation on the joint system $A_{1}A_{2}$. A simple global operation consists in letting systems $A_{1}$ and $A_{2}$ interact. In general, as a result of direct interactions, systems $A_{1}$
and $A_{2}$  become quantum correlated. On the other hand, entanglement can also be transferred from a third system $B_{1}B_{2}$ to $A_{1}A_{2}$. If systems $B_{1}$ and $B_{2}$ are entangled, then one can apply local operations on the pairs of systems $(A_{1}B_{1})$ and $(A_{2}B_{2})$. As a result of these operations it is possible to transfer the entanglement, originally in $B_{1}B_{2}$, to the joint system $A_{1}A_{2}$ (see Fig.{\ref{F:qubit-qubit}}). This approach is useful in the case where $A_{1}$ and $A_{2}$ represent two distant systems or when they interact weakly (or do not interact at all). The entanglement transfer from flying qubits to localized qubits  has been extensively studied (see \cite{Paris} and references therein).
 Entanglement transfer from two qubit systems to two qubit systems was investigated in \cite{Lee}.
  Moreover, in Fig.\ref{F:qubit-qubit}, the systems  $A_{i},B_{i}$, $(i=1,2)$ may represent different degrees of a freedom of single particle, such as spin and momentum. In fact, in \cite{Adami}, it was shown that the entanglement between the momenta of two different particles can be transferred to the spins of the particles under Lorentz transformations.
In the last few years,  entanglement transfer from many body systems and relativistic quantum field has been considered. In \cite{Vedral}, a scheme was proposed to extract entanglement from a quantum gas to a pair of qubits via local interactions. Also, it was shown in \cite{Reznik} that two qubits interacting locally with a quantum field can become entangled even when the qubits remain in causally disconnected regions throughout the whole interaction process. The previously described scenarios are depicted in Fig.\ref{F:qubit-background} where systems $A_{1}$ and $A_{2}$ are coupled to a common system B.

 In the present paper we investigate the entanglement transfer scheme illustrated in Fig.\ref{F:qubit-background} where $A_{1}$ and $A_{2}$ represent a pair of two level systems (qubits). Here, we assume that the group of operators involved in the interaction between $A_{1}$ and $B$  commute with the operators coupling $A_{2}$ to $B$. By doing so, we  mimic the entanglement transfer scheme described in Fig.\ref{F:qubit-qubit}. The difference between both cases lies in the fact that the Hilbert space $\mathcal{H}_{B}$ of system B may not have the structure $\mathcal{H}_{B_{1}}\otimes \mathcal{H}_{B_{2}}$. This point requires further explanation; for example, when systems $A_{1}$, $A_{2}$ are coupled to different segments of a chain of coupled harmonic oscillators, then one can identify the Hilbert spaces $\mathcal{H}_{B_{1}}$ and $\mathcal{H}_{B_{2}}$. However, when $A_{1}$ and $A_{2}$ couple to a quantum field vial local field operators then it is not clear what should be taken as the subsystems $B_{1}$ and $B_{2}$. On the other hand,  since operators acting on different Hilbert spaces commute, it is clear that the situation portrayed in Fig.\ref{F:qubit-qubit}, is a particular case of the scheme we investigate. \\
  The paper is organized as follows. In section (\ref{S:Basic}) we present some necessary conditions that the operators coupling the systems must satisfy in order to allow entanglement transfer from B to $A_{1}A_{2}$. Also, we derive an expression for the two qubit reduced density matrix. In section (\ref{S:Operators}), we introduce a special class of Hamiltonians describing the interaction between qubits and bosonic systems. For this class of  Hamiltonians (being  a generalization of the Jaynes-Cummings Hamiltonian), the reduced density matrix of the qubits assumes a particulary simple form. In section (\ref{S:Expansion}) we expand the entanglement measure (negativity) in terms of the coupling strength. We express the first nonvanishing contribution to this expansion in terms of 2-point and 4-point correlation functions involving the operators acting on system B. We study the entanglement of qubits  in the weak coupling approximation (or equivalently, for short interaction times), for different N-particle excitations of the bosonic system. We also investigate the entanglement of the qubits in the case where system B is in a mixed state of the form $\rho_{B}=\sum_{N}p_{N}\ket{N}\bra{N}$. Finally, in sections (\ref{S:N}) and (\ref{S:Ndiff}) we compute the exact  density matrix for the qubits  and discuss their entanglement for different N-particle states and operators.

\section{Formulation of the Problem}
\label{S:Basic}
Consider the interaction between system B, with Hilbert space $\mathcal{H}_{B}$, and system $A_{1}A_{2}$, with Hilbert space $\mathcal{H}_{A_{1}}\otimes \mathcal{H}_{A_{2}}$. This interaction induces the following global operation on the system $BA_{1}A_{2}$:
\beq \rho \rightarrow \mU \rho\mU^{\dagger}. \eeq
Assume that initially system {A} is in a separable state $\rho_{A}=\sum_{n}p_{n}\rho_{A_{1},n}\otimes \rho_{A_{2},n}$ while system B is in the state $\rho_{B}$. In addition, we assume that systems A and B are uncorrelated i.e.  $\rho=\rho_{A}\otimes \rho_{B}$.
After the interaction, the state of A is described by the reduced density matrix
\beq
{\label{redmat}}
\rho^{A}=\textrm{Tr}_{B}(\mU \rho_{A}\otimes \rho_{B}\mU^{\dagger})
\eeq
obtained by tracing out the degrees of freedom corresponding to system B. Making use of the spectral decomposition $\rho_{B}=\sum_{l}\lambda_{B,l}\ket{\lambda_{B,l}}\bra{\lambda_{B,l}}$ and choosing an orthonormal basis $\{\ket{\Phi_{B,j}}\}$ for $\mathcal{H}_{B}$,
one can write the operator-sum representation \cite{Kraus} of  the operation $\rho_{A}\rightarrow \rho^{A}$
\beq \rho^{A}=\sum_{j,l}\mathcal{A}_{j,l} \rho_{A} {{\mathcal{A}_{j,l}}^{\dagger}}. \eeq
The above expression shows that the positivity of the density matrix $\rho_{A}$ is preserved by the operation $\rho_{A}\rightarrow \rho^{A}$. The operators ${\mathcal{A}}_{j,l}$  are given by \beq
\label{Kraus}
\mathcal{A}_{j,l}=\sqrt{\lambda_{B,l}}\bra{\Phi_{B,j}}\mU \ket{\lambda_{B,l}}
\eeq
and they satisfy the relation $\sum_{j,l} \mathcal{A}_{j,l}{\mathcal{A}_{j,l}}^{\dagger}=\textrm{I}_{A}$ which guarantees that $\textrm{Tr}(\rho^{A})=1$. Let  $\mU=e^{-i \mH t}$ where $\mH$ is  a Hermitian operator of the form
 \beq
 \label{hamiltonian}
 \mH=\sum_{k} {\mA}_{1,k}\otimes {\mB}_{1,k}+ \sum_{l}{\mA}_{2,l}\otimes {\mB}_{2,l}.
 \eeq

     Here, ${\mA}_{i,k}$ and ${\mB}_{i,k}$ are Hermitian operators acting on $\mathcal{H}_{A_{i}}$, $i=(1,2)$ and $\mathcal{H}_{B}$, respectively. In addition, we assume that  $[{\mB}_{1,k}, {\mB}_{2,l}]=0$ which implies that the evolution operator factorizes as $\mU= \mU_{1}\cdot\mU_{2}$  where $\mU_{i}=e^{-it \sum_{k} {\mA}_{i,k}\otimes {\mB}_{i,k}}$,  $i=(1,2)$. It turns out that if all the operators coupled to one of the systems, say $A_{1}$,  commute i.e.
  \beq
  [{\mB}_{1,k},{\mB}_{1,k'}]=0
  \eeq
   then the state $\rho^{A}$ will remain separable. In order to prove this fact, we choose  $\{\ket{\Phi_{j}}\}$ in expression (\ref{Kraus})  to be a basis in which all the operators $\mB_{1,k}$ are diagonal, that is, $\mB_{1,k}\ket{\Phi_{B,j}}=b_{1,kj}\ket{\Phi_{j}}$. Then, the operators $A_{j,l}$ assume the form
  \beq
 \label{factor}
  \mathcal{A}_{j,l}={\mU_{1,j}}\otimes \mathcal{A}_{2,j,l}
  \eeq
   with $\mU_{1,j}=e^{-it \sum_{k}b_{1,kj}\mA_{1,k}}$ and $\mathcal{A}_{2,j,l}=\sqrt{\lambda_{B,l}}\braket{\Phi_{B,j}|\mU_{2}|{\lambda_{B,l}}}$. It is clear that operators of the form (\ref{factor})  map a separable state into another separable state. For example, if  $\rho_{A}=\rho_{A_{1}}\otimes \rho_{A_{2}}$ then using (\ref{Kraus}) one obtains
\beq
 \rho_{A_{1}}\otimes \rho_{A_{2}} \rightarrow \sum_{j,l} \textrm{Tr}_{A_{2}}(\mathcal{A}_{2,j,l}\rho_{A_{2}}{\mathcal{A}_{2,j,l}}^{\dagger})\mU_{1,j}\rho_{A_{1}}{\mU_{1,j}}^{\dagger}
 \frac{\mathcal{A}_{2,j,l}\rho_{A_{2}}{\mathcal{A}_{2,j,l}}^{\dagger}} { \textrm{Tr}_{A_{2}}(\mathcal{A}_{2,j,l}\rho_{A_{2}}{\mathcal{A}_{2,j,l}}^{\dagger})}
 \eeq
which is a convex sum of density matrices i.e.  $\rho^{A}=\sum_{j,l} p_{j,l}\rho_{A_{1},jl}\otimes \rho_{A_{2},jl}$, with  $\sum_{j,l}p_{j,l}=1$. Therefore, in order to entangle systems $A_{1}$ and $A_{2}$ each group of operators $\{\mB_{1,k}\}$ and $\{\mB_{2,k}\}$ in (\ref{hamiltonian}) must contain at least one pair of noncommuting operators. In particular, this fact rules out  operations of the form $\mU_{i}=e^{-it \mA_{i} \otimes \mB_{i}}$.  Notice, however, that this statement holds true for time independent  Hamiltonians of the form $(\ref{hamiltonian})$. If one incorporates the free evolution of systems A and B, then, in general, the time evolution operator $\mU(t)$ will contain noncommuting operators. In what follows, we will neglect the free evolution of the systems A and B. Furthermore, we will consider the situation where $A_{1}$  and $A_{2}$ are two-level systems (qubits) whereas B is a  bosonic system.\\
 In principle, one can obtain the reduced density matrix $\rho^{A}$ from Kraus representation(\ref{Kraus}). However, for our purposes it is  convenient to write  equation (\ref{redmat}) as
\beq
\label{fundamental}
\bra{a}\rho^{A}\ket{a'}=\textrm{Tr}_{B}(\rho_{B}\bra{\phi_{A}}\mU^{\dagger}\ket{a'}\bra{a}\mU\ket{\phi_{A}})
\eeq
which explicitly shows that the matrix elements of $\rho^{A}$ are  given by expectation values of operators acting on system {\bf{B}}. Since the unitary operator $\mU$ factorizes i.e. $\mU=\mU_{1}\cdot\mU_{2}$, one can define the operators $\mK_{i}\equiv \braket{0|\mU_{i}|0}$ and $\mN_{i}\equiv\braket{1|\mU_{i}|0}$ for (i=1,2). Assuming that the qubits are initially in the separable state $\phi_{A}=\ket{0,0}$  and choosing the basis $\{\ket{a_{1}}=\ket{0,0},  \ket{a_{2}}=\ket{0,1}, \ket{a_{3}}=\ket{1,0}, \ket{a_{4}}=\ket{1,1}\}$ for $\mathcal{H}_{A}=\mathcal{H}_{A_{1}}\otimes \mathcal{H}_{A_{2}}$, one can write the reduced density matrix (\ref{fundamental}) as

\begin{equation}
\label{bigmatrix}
\rho^{A}=\left( \begin{array}{cccc}
\braket{{\mK_{1}}^{\dagger}\mK_{1} {\mK_{2}}^{\dagger}\mK_{2}}&\braket{{\mK_{1}}^{\dagger}\mK_{1}{\mN_{2}}^{\dagger}\mK_{2}}&\braket{{\mN_{1}}^{\dagger}\mK_{1}
{\mK_{2}}^{\dagger}\mK_{2}}&\braket{{\mN_{1}}^{\dagger}\mK_{1}{\mN_{2}}^{\dagger}\mK_{2}}\\
\braket{{\mK_{1}}^{\dagger}\mK_{1}{\mK_{2}}^{\dagger}\mN_{2}}
&\braket{{\mK_{1}}^{\dagger}\mK_{1}{\mN_{2}}^{\dagger}\mN_{2}}&\braket{{\mN_{1}}^{\dagger}\mK_{1}{\mK_{2}}^{\dagger}\mN_{2}}
&\braket{{\mN_{1}}^{\dagger}\mK_{1}{\mN_{2}}^{\dagger}\mN_{2}}\\
\braket{{\mK_{1}}^{\dagger}\mN_{1}{\mK_{2}}^{\dagger}\mK_{2}}&
\braket{{{\mK_{1}}^{\dagger}\mN_{1}{\mN_{2}}^{\dagger}\mK_{2}}}&
\braket{{\mN_{1}}^{\dagger}\mN_{1}{\mK_{2}}^{\dagger}\mK_{2}}&\braket{{\mN_{1}}^{\dagger}\mN_{1}{\mN_{2}}^{\dagger}\mK_{2}}\\
\braket{{\mK_{1}}^{\dagger}\mN_{1}{\mK_{2}}^{\dagger}\mN_{2}}&\braket{{\mK_{1}}^{\dagger}\mN_{1}{\mN_{2}}^{\dagger}\mN_{2}}&
\braket{{\mN_{1}}^{\dagger}\mN_{1}{\mK_{2}}^{\dagger}\mN_{2}}&\braket{{\mN_{1}}^{\dagger}\mN_{1}{\mN_{2}}^{\dagger}\mN_{2}}
\end{array} \right)
\end{equation}
 where we used the notation $\braket{\hat{\mB}}=\textrm{Tr}_{B}(\rho_{B}\hat{\mB})$. Notice that the operators $\{\mN_{i},{\mN_{i}}^{\dagger},\mK_{i}, {\mK_{i}}^{\dagger}\}$ satisfy the relations
\beq
{\mK_{i}}^{\dagger}\mK_{i}+{\mN_{i}}^{\dagger}\mN_{i}=\mI_{B},\quad (i=1,2) \quad \Rightarrow \textrm{Tr}(\rho^{A})=1.
\eeq
 Thus, from the matrix (\ref{bigmatrix}) we see that all the properties of state $\rho^{A}$ (in particular, separability) depend on the interplay of the different correlation functions of the operators $\{\mN_{i},{\mN_{i}}^{\dagger},\mK_{i}, {\mK_{i}}^{\dagger}\}$.
\section{The Interaction Model \label{S:Operators}}
The most general form of the interaction between  qubit $A_{i}$ and the bosonic system is given by the Hamiltonian $\mH_{i}=\mathbb{I}_{i}\otimes \mB_{0,i}+\sigma_{x,i}\otimes\mB_{x,i}+\sigma_{y,i}\otimes\mB_{y,i}+\sigma_{z,i}\otimes\mB_{z,i}$
where $\mB_{k,i}$ are Hermitian operators. In this paper we  restrict our discussion to the case where $\mB_{0,i}=\mB_{z,i}=0$, $i=(1,2)$. Under this assumption, the Hamiltonian of the system can be written as $\mH=\mH_{1}+\mH_{2}$ where
\beq
\label{special}
\mathbb{H}_{i}=\begin{pmatrix} 0 & {\mF_{i}}^{\dagger}\\ \mF_{i}&0\\ \end{pmatrix}, \quad i=(1,2).\end{equation}
The evolution operator factorizes (i.e. $\mU=\mU_{1}\cdot\mU_{2}$) if the set of operators coupled to $A_{1}$ commute with those coupled to $A_{2}$. Consequently, we impose the following conditions:
 \beq
 \label{commuting}
 [\mF_{1},\mF_{2}]=[\mF_{1},{\mF_{2}}^{\dagger}]=[{\mF_{1}}^{\dagger},{\mF_{2}}^{\dagger}]=0.
\eeq
Now, one can express the operators $\mK_{i}\equiv\braket{0|\mU_{i}|0}$ and $\mN_{i}\equiv\braket{1|\mU_{i}|0}$ in terms of $\mF_{i}$ and ${\mF_{i}}^{\dagger}$. From (\ref{special}), one obtains
\begin{eqnarray}
 {\label{series1}}
 &&\mK_{i}= \bra{0}\mU_{i}\ket{0}=\sum_{k=0}^{\infty}\frac{(-1)^{k}}{(2k)!}t^{2k}({\mF_{i}}^{\dagger}\mF_{i})^{k}=
 \cos(\sqrt{{\mF_{i}}^{\dagger}\mF_{i}}t) \\
 {\label{series2}}
 &&\mN_{i}=\bra{1}\mU_{i}\ket{0}=(-i)\sum_{k=0}^{\infty}\frac{(-1)^{k}}
 {(2k+1)!}t^{2k+1}\mF_{i}({\mF_{i}}^{\dagger}\mF_{i})^{k}=-i \mF_{i}\frac{\sin(\sqrt{{\mF_{i}}^{\dagger}\mF_{i}}t)}{\sqrt{{\mF_{i}}^{\dagger}\mF_{i}}}.\\ \nonumber
\end{eqnarray}
 Since B is a bosonic system, one may write the operators  $\mF_{i}$ in their second quantization form. We will restrict our discussion to the class of operators expressible as
  \beq
  \label{class}
   F(a,a^{\dagger})=p(a^{\dagger}a)a^{n}+q(a^{\dagger}a){a^{\dagger}}^{m}
   \eeq
   where  $p(a^{\dagger}a)$ and $q(a^{\dagger}a)$ are functions of $a^{\dagger}a$ and $n,m \geq 1$. Notice that while the  above class includes field operators it does not include 1-body operators. If the eigenstates of the density matrix $\rho_{B}$ are also eigenstates of the particle number operator $ \hat{N}={\sum}_{k} a^{\dagger}(\phi_{B,k})a(\phi_{B,k})$ then from (\ref{bigmatrix}), ({\ref{series1}}) and (\ref{series2}) we see that some of the matrix elements of $\rho^{A}$ vanish. In fact, under these assumptions $\rho^{A}$  takes the form
\begin{equation}
 \label{simplematrix}
 \rho^{A}=\left( \begin{array}{cccc}
\rho_{11}&0&0&\rho_{14}\\
0&\rho_{22}&\rho_{23}&0\\
0&\rho_{23}^{*}&\rho_{33}&0\\
\rho_{14}^{*}&0&0&\rho_{44}\\
\end{array} \right).
\end{equation}

The eigenvalues of $\rho^{A}$ are given by
 $\lambda_{1,\pm}=\frac{1}{2} (\rho_{11}+\rho_{44} \pm \sqrt{(\rho_{11}+\rho_{44})^2-4(\rho_{11}\rho_{44}-|\rho_{14}|^2})$ and
 $\lambda_{2,\pm}=\frac{1}{2}(\rho_{22}+\rho_{33} \pm \sqrt{(\rho_{22}+\rho_{33})^2-4(\rho_{22}\rho_{33}-|\rho_{23}|^2}).$
  Making use of Schwarz inequality ($|\braket{x|y}|< ||x||\cdot ||y||$), one easily proves the relations
   $\rho_{11}\rho_{44}\geq|\rho_{14}|^2$ and
  $\rho_{22}\rho_{33}\geq|\rho_{23}|^2$ which ensure the positivity of the density matrix $\rho_{A}.$
   On the other hand, the partial transpose \cite{Peres} of $\rho^{A}$ is defined as
   \[\bra{a_{1},a_{2}}\rho^{T_{A_{1}}}\ket{a_{1}',a_{2}'}=\bra{a_{1}',a_{2}}\rho^{A}\ket{a_{1},a_{2}'}\]
 and its matrix representation is
 \begin{equation}
 \label{PT}
   \rho^{T_{A_{1}}}=\begin{pmatrix}
   \rho_{11}&0&0&\rho_{23}^{*}\\
   0&\rho_{22}&\rho_{14}^{*}&0\\
   0&\rho_{14}&\rho_{33}&0\\
   \rho_{23}&0&0&\rho_{44}\\
   \end{pmatrix}.
   \end{equation}
 Notice that $\rho^{T_{A_{1}}}$ can be obtained from $\rho^{A}$   replacing $\rho_{23}$ by $\rho_{14}^{*}$ and $\rho_{14}$ by $\rho_{23}^{*}$. Hence the matrix $\rho^{T_{A_{1}}}$ will have a negative eigenvalue if either
 \beq
 {\label{inequalities}}
 n_{23}\equiv|\rho_{23}|^2-\rho_{11}\rho_{44} >0\quad \textrm{or} \quad n_{14}\equiv|\rho_{14}|^2 -\rho_{22}\rho_{33}>0.
\eeq
 If one of the above inequalities is fulfilled, then according to the PPT criterion \cite{Hodorecki}, the two qubit system  ${{A_{1}A_{2}}}$ will be in a nonseparable (entangled) state.
 Furthermore, in this particular case, one can show that $\rho^{A}>0$ implies that $n_{23}$ and $n_{14}$ cannot be positive at the same time. For example, suppose that $n_{23}>0$, then one can write the inequalities
 \[ \rho_{22}\rho_{33}\geq|\rho_{23}|^2>\rho_{11}\rho_{44}\geq|\rho_{14}|^2\]
which imply $n_{14}<0$. Therefore the partial transpose $\rho^{T_{A_{1}}}$ can only have one negative eigenvalue. In fact, it can be proved that in $2 \times 2$ dimensions, the partial transpose of any density matrix can have at most one negative eigenvalue \cite{Sanpera}.\\
In the previous section, we assumed that the initial state of system ${{A_{1}A_{2}}}$ was $\ket{\phi_{A}}=\ket{a_{1}}=\ket{0,0}$. Consequently, we wrote down the  matrix (\ref{bigmatrix}) representing the final state of the qubits. However, if system ${{A_{1}A_{2}}}$ is initially in a different state then, obviously, the matrix elements of $\rho^{A}$ will be different from those in expression (\ref{bigmatrix}). Let us consider, for example,  the case $\ket{\phi_{A}}=\ket{1,1}=\sigma_{x}\otimes \sigma_{x}\ket{0,0}.$
 Now, equation (\ref{fundamental}) can be written as
 \beq
 \bra{a}\rho^{A}\ket{a'}=\textrm{Tr}_{B}(\rho_{B}\bra{a_{1}}\sigma_{x}^{\dagger}\mU_{1}^{\dagger}\sigma_{x}\otimes
 \sigma_{x}^{\dagger}\mU_{2}^{\dagger}\sigma_{x}\ket{\bar{a'}}\bra{\bar{a}}\sigma_{x}^{\dagger}\mU_{1}
 \sigma_{x}\otimes\sigma_{x}^{\dagger}\mU_{2}\sigma_{x}\ket{a_{1}})
 \eeq
 where $\ket{\bar{a}}$ is obtained from $\ket{a}$ by flipping both qubits, i.e. $\ket{\bar{a}}=\sigma_{x}\otimes \sigma_{x}\ket{a}$.  Since
  \beq
\sigma_{x}^{\dagger}\mathbb{H}_{k}\sigma_{x}=\begin{pmatrix} 0 & \mF_{k}\\ {\mF_{k}}^{\dagger}&0\\
\end{pmatrix},
\eeq
  it is easy to conclude that if initially system {\bf{A}} was known to be in the state $\ket{a_{4}}=\ket{1,1} \in \mathcal{H}_{A_{1}}\otimes \mathcal{H}_{A_{2}}$ then the final density matrix $\rho^{A}$ is given by
  \begin{equation}
  {\rho^{A}}(\ket{a_{4}}, \mU) =\begin{pmatrix}
   \tilde{\rho}_{44}&0&0&\tilde{{\rho}}_{41}\\
   0&\tilde{{\rho}}_{33}&\tilde{{\rho}}_{32}&0\\
   0&\tilde{\rho}_{32}^{*}&\tilde{{\rho}}_{22}&0\\
   \tilde{\rho}_{14}&0&0&\tilde{\rho}_{11}\\
   \end{pmatrix}
   \end{equation}
  where the matrix elements  $\tilde{\rho}_{ij}$ are determined by means of the expressions
  (\ref{bigmatrix}),(\ref{series1}),(\ref{series2})  with the operators $\mF_{k}$  replaced by
  the operators $\mF_{k}^{\dagger}$ and vice-versa. Treating  $\rho^{A}$ as a function of the initial state $\ket{\phi_{A}}$ and the operator $\mU$, one can summarize the previous discussion as
  \beq
  \rho^{A}(\ket{a_{4}},\mU(\mF_{k}))=V\rho^{A}(\ket{a_{1}},\mU(\mF_{k}^{\dagger}))V^{\dagger}
  \eeq
    where $V=\sigma_{x}\otimes \sigma_{x}=\ket{a_{1}}\bra{a_{4}}+\ket{a_{4}}\bra{a_{1}}+\ket{a_{2}}\bra{a_{3}}+\ket{a_{3}}\bra{a_{2}}$ is the unitary transformation corresponding to the basis permutation
    $(1,2,3,4) \rightarrow (4,3,2,1)$.
    Similarly, one can write
    \beq
    \rho^{A}(\ket{a_{2}}, \mU(\mF_{1},\mF_{2}))=V \rho^{A}(\ket{a_{1}},\mU(\mF_{1},\mF_{2}^{\dagger}))V^{\dagger}
   \eeq
   with $V=I\otimes \sigma_{x}$.

\section{Series Expansion for $\mathcal{N}(\rho^{A})$ }
\label{S:Expansion}
 In general, for a given set of operators $\{\mF_{1},\mF_{2}\}$, it may not be possible to determine the matrix elements of $\rho^{A}$ in closed form. On the other hand, it is interesting to explore the different terms in the series expansion of $n_{23}$, $n_{14}$ (see \ref{inequalities}) in terms of the operators $\{\mF_{i},\mF_{i}^{\dagger}\}$. From the general expression $(\ref{bigmatrix})$ for the reduced density matrix, we expect this series expansion to contain correlation functions of the operators  $\{\mF_{i},{\mF_{2}}^{\dagger}\}$ . Introducing a coupling constant, i.e. $\mF_{i} \rightarrow g_{i}\mF_{i}$, one obtains a series expansion of the form
 \beq
 \label{exp}
{n}(g_{1},g_{2})= n_{1}g_{1}^{2}g_{2}^{2}+n_{2}(g_{1}^{2}g_{2}^{4}+g_{1}^{4}g_{2}^{2})+\ldots
 \eeq
 where $n$ denotes either  $n_{23}$ or $n_{14}$.  The above form can be justified as follows. Each term of the series expansion should be symmetric in $g_{1}$ and $g_{2}$. Terms containing odd powers of $g_{i}$, e.g. ${g_{1}g_{2}^{3}+g_{1}^{3}g_{2}}$,  should not be present since then one could change the sign of $n$ by simply changing the signs of $g_{1}$ or $g_{2}$. Clearly the entanglement in $A_{1}A_{2}$ should not depend on the sign of $g_{1}$ or $g_{2}$. Finally, if we switch off one of the interactions ($g_{1}$=0 or  $g_{2}=0$) then both  $n_{23}$ and $n_{14}$ vanish (see expressions (\ref{inequalities})). This rules out terms of the form $g_{1}^{4}+g_{2}^{4}$. Now, we proceed to express the first nonvanishing contribution to the expansion (\ref{exp}) in terms of $\mF_{i}$ and ${\mF_{i}}^{\dagger}$. Making use of the series expansion (\ref{series1}) and (\ref{series2}) for $\mK_{i}$ and $\mN_{i}$, one writes  $\mK_{i}=\mathbb{I}-\frac{1}{2}({\mF_{i}}^{\dagger}{\mF_{i}})t^2+\frac{1}{4!}({\mF_{i}}^{\dagger}{\mF_{i}})^2t^4\ldots $
and $\mN_{i}=-i\mF_{i}t(\mathbb{I}-\frac{1}{3!}{\mF_{i}}^{\dagger}{\mF_{i}}t^2)+\ldots $. From (\ref{bigmatrix}) one obtains the following series expansion for the diagonal matrix elements $\rho_{kk}$:
\begin{align}
\rho_{11}&=\mathbb{I}-\braket{\mP_{1}}t^2-\braket{\mP_{2}}t^2+\frac{1}{3}(\braket{\mP_{1}^2}t^4+\braket{\mP_{2}^2})t^4+\braket{\mP_{1}\mP_{2}}t^4,
& \rho_{22}=\braket{\mP_{2}}t^2-\frac{1}{3}\braket{\mP_{2}^2}t^4-\braket{\mP_{1}\mP_{2}}t^4 , \\
\rho_{33}&=\braket{\mP_{1}}t^2-\frac{1}{3}\braket{\mP_{1}^2}t^4-\braket{\mP_{1}\mP_{2}}t^4, &  \rho_{44}=\braket{\mP_{1}\mP_{2}}t^4
\end{align}
where $\mP_{i}\equiv ({\mF_{i}}^{\dagger} \mF_{i})$.  Similarly, one expands the $\rho_{23}$ and $\rho_{14}$ in series of $\{\mF_{i},{\mF_{i}}^{\dagger}\}$ obtaining the following expressions:
\begin{eqnarray}
\label{offdiagonal1}
\rho_{23}&=&\braket{{\mF_{1}}^{\dagger}{\mF_{2}}}t^2-\frac{1}{2}\braket{{\mF_{1}}^{\dagger}
{\mF_{2}}^{\dagger}{{\mF_{2}}}^2}t^4-
\frac{1}{2}\braket{{{\mF_{1}}^{\dagger}}^2 {\mF_{1}}\mF_{2}}t^4-\frac{1}{3!}\braket{{\mF_{1}}^{\dagger}{\mF_{1}}{\mF_{1}}^{\dagger}{\mF_{2}}}t^4 \nonumber \\
&&-\frac{1}{3!}\braket{{\mF_{1}}^{\dagger}{\mF_{2}}{\mF_{2}}^{\dagger}{\mF_{2}}}t^4\\
\label{offdiagonal2}
\rho_{14}&=&-\braket{{\mF_{1}}^{\dagger}{\mF_{2}}^{\dagger}}t^2+\frac{1}{2}\braket{{\mF_{1}}^{\dagger}
{{\mF_{2}}^{\dagger}}^2 \mF_{2}}t^4+
\frac{1}{2}\braket{{{\mF_{1}}^{\dagger}}^{2}\mF_{1}{\mF_{2}}^{\dagger}}t^4+\frac{1}{3!}\braket{{\mF_{1}}^{\dagger}{\mF_{2}}^{\dagger}
\mF_{2}{\mF_{2}}^{\dagger}}t^4  \nonumber \\
&&+\frac{1}{3!}\braket{{\mF_{1}}^{\dagger}\mF_{1}{\mF_{1}}^{\dagger}{\mF_{2}}^{\dagger}}t^4.\\ \nonumber
\end{eqnarray}
Thus up to fourth order one has
\begin{eqnarray} \rho_{11}\rho_{44}&=& \braket{\mP_{1}\mP_{2}}t^4=\braket{{\mF_{1}}^{\dagger}\mF_{1}{\mF_{2}}^{\dagger}\mF_{2}}t^4=
\braket{(\mF_{1}\mF_{2})^{\dagger}\mF_{1}\mF_{2}}t^4 \geq 0
\\
 |\rho_{14}|^2&=&|\braket{{\mF_{1}}^{\dagger}{\mF_{2}}^{\dagger}}|^2t^4 \\
 \rho_{22}\rho_{33}&=&\braket{\mP_{1}}\braket{\mP_{2}}t^4=\braket{{\mF_{1}}^{\dagger}\mF_{1}}\braket{{\mF_{2}}^{\dagger}\mF_{2}}t^4 \geq 0\\
|\rho_{23}|^2&=&|\braket{{\mF_{1}}^{\dagger}\mF_{2}}|^{2}t^4.\\ \nonumber
 \end{eqnarray}

 For consistency, one can check the density matrix is positive. In fact, the identity $\braket{\mA\mA^{\dagger}}\geq |\braket{\mA}|^2 $ with $\mA={\mF_{1}}^{\dagger}{\mF_{2}}^{\dagger}$ implies that $\rho_{11}\rho_{44} \geq |\rho_{14}|^2$ while the inequality $\rho_{22}\rho_{33}\geq |\rho_{23}|^2$ follows from Schwarz inequality. Clearly, the  partial transpose $\rho^{T_{A_{1}}}$ may be negative. In fact, substituting the above approximations for $\rho^{A}$ in  the expressions (\ref{inequalities}) for  $n_{23}$ and $n_{14}$ one obtains:
\begin{eqnarray}
\label{ns}
&&n_{23}=t^4(|\braket{{\mF_{1}}^{\dagger}\mF_{2}}|^2-\braket{{\mF_{1}}^{\dagger}\mF_{1}
{\mF_{2}}^{\dagger}\mF_{2}})\\
\label{ns1}
&&n_{14}=t^4(|\braket{{\mF_{1}}^{\dagger}{\mF_{2}}^{\dagger}}|^2-\braket{{\mF_{1}}^{\dagger}\mF_{1}}\braket{{\mF_{2}}^{\dagger}\mF_{2}}).\\
\nonumber
\end{eqnarray}
 For small values of t, the above expressions can be used to detect the presence of entanglement in the system $\bf{A_{1}A_{2}}$. If one of the above quantities ($n_{23}$ or $n_{14}$) is positive then the state of qubits $A_{1}$ and $A_{2}$ is nonseparable. Notice that if we choose the operators $\mF_{i}$ to be field operators acting on B, then $n_{23}$ and $n_{14}$ will contain 2-point and 4-point correlation functions of these operators. On the other hand, if the operators $\{\mF_{1},\mF_{2}\}$ are normal (i.e. $[\mF_{i},{\mF_{i}}^{\dagger}]=0$) then $n_{23}$ and $n_{14}$ are be negative. This is a reflection of the fact that in order to entangle system $A_{1}$ with system $A_{2}$ we must have non commuting operators. In the particular case where the operators $\mF_{i}$ are linear combinations of creation and annihilation operators we have
\beq
\label{linear}
[\mF_{i},{\mF_{i}}^{\dagger}]=c_{i} \quad \textrm{for} \quad (i=1,2),
\eeq
where $c_{i}$ is a c-number. In this case one can show that $n_{23}$ and $n_{14}$ are negative if either $c_{1}<0$ or $c_{2}<0$. For example, let $c_{2}<0$, then Schwarz inequality leads to
\begin{eqnarray}
n_{14}&=&t^4(|\braket{{\mF_{1}}^{\dagger}{{\mF_{2}}}^{\dagger}}|^2-\braket{{\mF_{1}}^{\dagger}{\mF_{1}}}\braket{{\mF_{2}}{\mF_{2}}^{\dagger}}+
c_{2}\braket{{\mF_{1}}^{\dagger}{\mF_{1}}})\leq c_{2}t^4
\braket{{\mF_{1}}^{\dagger}{\mF_{1}}} \leq 0 \\
 \label{last}
 n_{23}&=&t^4(|\braket{{\mF_{1}}^{\dagger}{\mF_{2}}}|^2-\braket{({\mF_{1}}^{\dagger}{\mF_{2}}) ({\mF_{1}}^{\dagger}{\mF_{2}})^{\dagger}}+c_{2}\braket{{\mF_{1}}^{\dagger}\mF_{1}})\leq c_{2}t^4\braket{{\mF_{1}}^{\dagger}{\mF_{1}}}\leq 0. \\ \nonumber
 \end{eqnarray}
The relation (\ref{last}) is obtained using the inequality
\beq\braket{\mA {\mA}^{\dagger}}\geq |\braket{\mA}|^2.\eeq

Finally, the entanglement measure we use in this paper is the negativity  $\mathcal{N}(\rho^{A})$
defined as twice the absolute value of the negative eigenvalue of $\rho^{T_{A_{1}}}$\cite{Werner}.  In our case, the eigenvalues of the partial transpose $\rho^{T_{A_{1}}}$ (\ref{PT}) are (up to fourth order in t):
\begin{eqnarray}
\lambda^{T_{A_{1}}}_{1,-}&=&\frac{1}{2} (\rho_{11}+\rho_{44} - \sqrt{(\rho_{11}+\rho_{44})^2+4 n_{23}})\approx -\frac{n_{23}}{\rho_{11}+\rho_{44}}\approx -n_{23}\\
\lambda^{T_{A_{1}}}_{2,-}&=&\frac{1}{2}(\rho_{22}+\rho_{33} - \sqrt{(\rho_{22}+\rho_{33})^2+4n_{14}})\approx -\frac{n_{14}}{\rho_{22}+\rho_{33}}. \\ \nonumber
\end{eqnarray}
 \subsection{Examples}
Utilizing expressions $(\ref{ns})$ and $(\ref{ns1})$  one can calculate the entanglement transferred from system B to the pair of qubits $A_{1}A_{2}$. In this subsection we compute the quantities $n_{23}$ and $n_{14}$ for interactions of the form (\ref{class}) i.e. $F_{i}(a,a^{\dagger})=f(a^{\dagger}a)a^{n}+g(a^{\dagger}a){a^{\dagger}}^{m}$ and statistical mixtures of N-particle states $\rho_{B}=\sum_{N}p_{N}\ket{N}\bra{N}$. Most of the examples that we present here can be computed exactly using (\ref{bigmatrix}). Nevertheless we believe that the first nonvanishing contribution to entanglement gives us some hints about the states and interactions that induce entanglement in system $A_{1}A_{2}$.  Moreover, from a technical point of view, expressions of the form  $\braket{{\mF_{1}}^{\dagger}{{\mF_{2}}}^{\dagger}}$, $\braket{{\mF_{1}}^{\dagger}\mF_{1}{\mF_{2}}^{\dagger}\mF_{2}} \ldots $
 can be easily computed using Wick Theorem \cite{Peskin}.

First, we consider the JC \cite{JC} describing a system of two two-level atoms interacting with a set of electromagnetic modes. Let the creation and annihilation operators for these modes be $a_{k}$ and $a_{k}^{\dagger}$. We assume that the Hamiltonian for the two atoms is of the form $\mH=\mH_{1}+\mH_{2}$ with $\mH_{i}$, $(i=1,2)$ given by
  \beq
  \label{JC}
  \mH_{i}=\sum_{k}g_{i,k}(\sigma_{+}a_{k}+ \sigma_{-}{a_{k}}^{\dagger}), \quad \sigma_{+}=\begin{pmatrix} 0 &0, \\
                                   1&0\\ \end{pmatrix}, \quad \sigma_{-}=\begin{pmatrix}  0& 1 \\
                                                                                          0 &0\\ \end{pmatrix}.
  \eeq
 In this representation, $\ket{0}=\begin{pmatrix} 1 \\ 0 \\ \end{pmatrix}$ denotes the ground state of the atom while $\ket{1}=\begin{pmatrix} 0\\1 \end{pmatrix}$ represents the excited state of the atom. Let  $\ket{k}$  denote the mode created by the operator $a^{\dagger}_{k}$. Introducing the state $\ket{\phi_{i}}=\sum_{k}
  \frac{g_{i,k}}{\sqrt{\sum_{k}g_{i,k}^{2}}}\ket{k}$ we rewrite the Hamiltonian (\ref{JC}) as
\beq
\mH_{i}=g_{i}(\sigma_{+}a(\phi_{i})+\sigma_{-}a^{\dagger}(\phi_{i}))=g_{i}\begin{pmatrix} 0 & a^{\dagger}(\phi_{i})\\

                                                                                    a(\phi_{i})&0 \\  \end{pmatrix} \eeq
where $g_{i}=\sqrt{\sum_{k} g_{i,k}^{2}} $.  This Hamiltonian is of the form  $\mathbb{H}_{i}=\begin{pmatrix}0 & {\mF_{i}}^{\dagger}\\ \mF_{i} &0 \\ \end{pmatrix}$ with $
\mF_{i}=g_{i}a^{\dagger}(\phi_{i})$. These operators satisfy $[\mF_{1}, \mF_{2}]=0$, $[\mF_{i},{\mF_{i}}^{\dagger}]=g_{i}^2$ and $[\mF_{1},{\mF_{2}}^{\dagger}]=g_{1}g_{2}\braket{\phi_{1}|\phi_{2}}.$
 If we assume that the states $\ket{\phi_{1}}$ and $\ket{\phi_{2}}$ are orthogonal, then relations (\ref{commuting}) will be satisfied and we can apply, the results from previous sections. From the physical point of view, the orthogonality of $\ket{\phi_{1}}$ and $\ket{\phi_{2}}$ could describe the situation in which two atoms are sensible to different modes i.e. they  absorb/emit particles in different modes. If initially the qubits are prepared in the state  $\ket{\phi_{A}}=\ket{0,0}$ while system $\bf{B}$  is in the state $\ket{\Phi_{B}}=\frac{1}{\sqrt{N!}}(a^{\dagger}(\phi_{B}))^{N}\ket{0}$  (representing N bosons occupying the same state $\ket{\phi_{B}}$) one easily finds that
\begin{eqnarray}
\label{mat1}
\braket{{\mF_{1}}^{\dagger}\mF_{2}}&=&g_{1}g_{2}\braket{{a_{1}}^{\dagger}a_{2}}= g_{1}g_{2}N\braket{\phi_{B}|\phi_{1}}\braket{\phi_{2}|\phi_{B}}  \\
\label{mat2}
 \braket{{\mF_{1}}^{\dagger}\mF_{1}{\mF_{2}}^{\dagger}\mF_{2}}&=&g_{1}^{2}g_{2}^{2}\braket{{a_{1}}^{\dagger}a_{1}{a_{2}}^{\dagger}a_{2}}=g_{1}^{2}g_{2}^{2}
N(N-1)|\braket{\phi_{1}|\phi_{B}}|^2 |\braket{\phi_{2}|\phi_{B}}|^2. \\ \nonumber
\end{eqnarray}
  Substituting these results in expressions (\ref{ns}) and (\ref{ns1}) one obtains:
\begin{eqnarray}
\label{23positive}
 n_{23}&=&Ng_{1}^{2} g_{2}^{2}t^{4}|\braket{\phi_{1}|\phi_{B}}|^2 |\braket{\phi_{2}|\phi_{B}}|^2 >0\\ \nonumber
 n_{14}&=&-N^{2}g_{1}^{2} g_{2}^{2} t^{4}|\braket{\phi_{1}|\phi_{B}}|^2 |\braket{\phi_{2}|\phi_{B}}|^2<0. \\ \nonumber
 \end{eqnarray}
  From the above expressions we conclude that system $A_{1}A_{2}$ is always entangled except for
   the cases where $u_{1}\equiv\braket{\phi_{1}|\phi_{B}}=0$ or $u_{2}\equiv\braket{\phi_{2}|\phi_{B}}=0$. One can also consider the situation where the initial state of system B is a mixed state of the form
    \beq
   \rho_{B}=\sum_{N=0}^{\infty}p_{N}\ket{N}\bra{N}, \quad  \quad \sum_{N=0}^{\infty} p_{N}=1,
   \eeq
 and $\ket{N}=\frac{1}{\sqrt{N!}}(a^{\dagger}(\phi_{B}))^{N}\ket{0}$.  As discussed in section (\ref{S:Operators}), all the expressions we have derived so far hold for the class of density matrices having eigenstates with fixed particle number. Thus, from equations (\ref{mat1}) and
 (\ref{mat2}), one obtains
 \beq
 n_{23}=g_{1}^{2}g_{2}^{2}t^{4}|u_{1}|^{2}|u_{2}|^{2}((\sum_{N=0}^{\infty}p_{N}N)^{2}-\sum_{N=0}^{\infty}p_{N}N(N-1))
 \eeq
 which implies that $\rho^{A}$ will be nonseparable for probability distributions $p_{N}$ satisfying
 \beq
  (\sum_{N}^{\infty}p_{N}N)^{2}>  \sum_{N}^{\infty}p_{N}N(N-1).
  \eeq
It is interesting to note that the above condition implies that the probability distribution $\{p_{N}\}$ must be sub-poissonian \cite{Mandel} .i.e. $ \bar{N}> \bar{N^{2}}-(\bar{N})^{2}=\sigma_{N}^2$.  An example of a distribution satisfying $\bar{N}> \sigma_{N}^2$ is the binomial distribution $p_{N}=\binom{M}{N}p^{N}(1-p)^{M-N}$ (for $N=0,1, \ldots M$) having $\bar{N}=Mp$ and $\sigma_{N}^{2}=Mp(1-p)$. In this particular case one obtains
\beq
n_{23}=g_{1}^{2}g_{2}^{2}t^{4}|u_{1}|^{2}|u_{2}|^{2}Mp^{2}>0.
\eeq
Finally, notice that a thermal-like state $\rho_{B}=(1-z)\sum_{N} z^{N} \ket{N}\bra{N}$ yields a separable state for ${A_{1}A_{2}}$.
  If the initial state of the two-level systems is $\ket{\phi_{A}}=\ket{1,1}$ then according to section (\ref{S:Operators}), the entanglement of $A_{1}A_{2}$ is determined by $n_{23}$ and $n_{14}$ with the operators $\mF_{i}$ being replaced by ${\mF_{i}}^{\dagger}$. As expected, in this case we obtain a separable state. In fact, if we denote $u_{i}=\braket{\phi_{i}|\phi_{B}}$, then we have \beq
|\braket{\mF_{1}{\mF_{2}}^{\dagger}}|^{2}-\braket{\mF_{1}{\mF_{1}}^{\dagger}\mF_{2}{\mF_{2}}^{\dagger}}=-g_{1}^{2} g_{2}^2N(|u_{1}|^2+|u_{2}|^2-|u_{1}|^2|u_{2}|^2+\frac{1}{N})<0.
\eeq

One can also consider a state of the form
\beq
\ket{\Phi_{B}}=
\frac{1}{\sqrt{N_{1}!N_{2}!}}(a^{\dagger}(\phi_{B_{1}}))^{N_{1}}(a^{\dagger}(\phi_{B_{2}}))^{N_{2}}\ket{0}\equiv\ket{N_{1}, N_{2}}
\eeq
describing the situation where $N_{1}$ bosons occupy the state $\ket{\phi_{{B_{1}}}}$ and $N_{2}$ bosons occupy the state  $\ket{\phi_{{B_{2}}}}$  orthogonal to $\ket{\phi_{B_{1}}}$, i.e. $\braket{\phi_{B_{1}}|\phi_{B_{2}}}=0$. Straightforwardly, one finds that
\begin{eqnarray}
a_{i}\Ket{\Phi_{B}}&=&\sqrt{N_{1}}u_{i,1}\ket{N_{1}-1,N_{2}}+\sqrt{N_{2}}u_{i,2}\ket{N_{1},N_{2}-1}\\
|\braket{{\mF_{1}}^{\dagger}\mF_{2}}|^2&=&g_{1}^{2}g_{2}^{2}|N_{1}{{u^{*}}_{1,1}}u_{2,1}+N_{2}{{u^{*}}_{1,2}}u_{2,2}|^2\\ \nonumber
\end{eqnarray}
where $u_{i,k}\equiv \braket{\phi_{i}|{\phi}_{B_{k}}}$. On the other hand, $\braket{{\mF_{1}}^{\dagger}\mF_{1}{\mF_{2}}^{\dagger}\mF_{2}}$ reads:
\beq
\braket{{\mF_{1}}^{\dagger}\mF_{1}{\mF_{2}}^{\dagger}\mF_{2}}
=g_{1}^{2}g_{2}^{2}(N_{1}(N_{1}-1)|u_{1,1}u_{2,1}|^2+N_{2}(N_{2}-1)|u_{2,2}u_{1,2}|^2+N_{1}N_{2}|u_{2,1}u_{1,2}+u_{2,2}u_{1,1}|^2).\eeq
Combining the above results one obtains
\beq
\label{N1N2}
n_{23}=g_{1}^{2}g_{2}^{2}t^{4}(N_{1}|u_{1,1} u_{2,1}|^2+N_{2}|u_{1,2}u_{2,2}|^2-N_{1}N_{2}(|u_{2,1}u_{1,2}|^2+|u_{2,2}u_{1,1}|^2)).
\eeq
Here, two things are worth mentioning. First of all, equation (\ref{N1N2}) indicates, that in this approximation,  no entanglement will be transferred to $A_{1}A_{2}$ for large values of $N_{1}$ and $N_{2}$. Also notice that when $N_{1}=N_{2}=1$, $|\braket{{\mF_{1}}^{\dagger}\mF_{2}}|^2=g_{1}^{2}g_{2}^{2}|N_{1}{{u^{*}}_{1,1}}u_{2,1}+N_{2}{{u^{*}}_{1,2}}u_{2,2}|^2$  vanishes when the $2\times 2$ matrix $u_{i,k}=\braket{\phi_{i}|{\phi}_{B_{k}}}$ is unitary implying $n_{23}<0$. This case will be studied in detail in section (\ref{S:Ndiff}).

\subsection{Algebraic Construction of Operators}

 One can also consider the entanglement induced in system  $A_{1}A_{2}$ when the operators $\mF_{i}$ are linear combinations of creation and annihilation operators. Let
\beq
\label{mix}
 \mF_{i}=g_{i}(a_{i}+|\beta_{i}|e^{i \theta_{i}}{a_{i}}^{\dagger}), \quad i=(1,2), \quad \textrm{with} \quad [a_{1},{a_{2}}^{\dagger}]=0 \quad \textrm{and} \quad |\beta_{i}|<1.
 \eeq
 If system B is in a 1-particle state $a_{B}^{\dagger}\ket{0}$, then
 \begin{eqnarray}
 \braket{{\mF_{1}}^{\dagger}\mF_{2}}&=&\braket{0|a_{B}{\mF_{1}}^{\dagger}\mF_{2}{a_{B}}^{\dagger}|0}=g_{1}g_{2}
 ({u_{1}}^{*}u_{2}+u_{1}{u_{2}}^{*}|\beta_{1}|\beta_{2}|e^{-i(\theta_{1}-\theta_{2})})\\
\braket{{\mF_{1}}^{\dagger}\mF_{1}{\mF_{2}}^{\dagger}\mF_{2}}&=&
\braket{0|a_{B}{\mF_{1}}^{\dagger}\mF_{1}{\mF_{2}}^{\dagger}\mF_{2}{a_{B}}^{\dagger}|0}\nonumber \\
&=&g_{1}^2 g_{2}^2(|\beta_{1}|^2|\beta_{2}|^2+|u_{1}|^2(1+|\beta_{1}|^2)|\beta_{2}|^2+|u_{2}|^2(1+|\beta_{2}|^2)|\beta_{1}|^2)
\end{eqnarray}
In the case where $u_{1}=u_{2}=\frac{1}{\sqrt{2}}$ and $\beta_{1}=\beta_{2}=\beta$ we have
  \beq
  \braket{{\mF_{1}}^{\dagger}\mF_{2}}-\braket{{\mF_{1}}^{\dagger}\mF_{1}{\mF_{2}}^{\dagger}\mF_{2}}>0 \quad \textrm{for} \quad |\beta|<|\beta_{max}|=\sqrt{\frac{2\sqrt{2}-1}{7}}=0.51
   \eeq
   and $n_{14}=t^{4}(\braket{{\mF_{1}}^{\dagger}{\mF_{2}}^{\dagger}}-\braket{\mF_{1}^{\dagger}\mF_{1}}\braket{{\mF_{2}}^{\dagger}\mF_{2}})<0$
  for $\beta \in (0,1).$ Thus, one can mix creation and annihilation operators as in equation (\ref{mix}), obtaining an entangled state for $A_{1}A_{2}$ for $|\beta|< |\beta_{max}|$.

   We conclude this section by constructing a set of operators inducing entanglement in system  $A_{1}A_{2}$  when B is the particle vacuum state $\ket{0}$. This problem was studied in \cite{Reznik} where it was shown using perturbation theory  that a two-level systems can become entangled after having locally interacted with a scalar field. Here, we assume an effective interaction between the qubits and system B of the form
  \beq
  \label{field}
  \mH_{i}=\begin{pmatrix} 0 & {\mF_{i}}^{\dagger}\\\mF_{i}& 0 \\  \end{pmatrix} \quad \textrm{with} \quad  \mF_{i}=g_{i}(a(\phi_{i})+\beta_{i}^{*}a^{\dagger}(\psi_{i})).
  \eeq

  We need to accommodate conditions (\ref{commuting}) into this picture. Notice that if $\braket{\phi_{1}|\phi_{2}}=\braket{\psi_{1}|\psi_{2}}=0 $ then $[\mF_{1},{\mF_{2}}^{\dagger}]=0.$ On the other hand,
$[\mF_{1},\mF_{2}]=0$ holds if we impose the condition
 \beq
 \beta_{1}\braket{\psi_{1}|\phi_{2}}=\beta_{2}\braket{\psi_{2}|\phi_{1}}.
 \eeq
  Now, we have
 \begin{eqnarray}
 \braket{N|{\mF_{1}}^{\dagger}{\mF_{2}}^{\dagger}|N}&=&g_{1}g_{2}(\beta_{1}\braket{\psi_{1}|\phi_{2}}+
 N(\braket{\phi_{B}|\phi_{1}}\beta_{2}\braket{\psi_{2}|\phi_{B}}+\braket{\phi_{B}|\phi_{2}}\beta_{1}\braket{\psi_{1}|\phi_{B}})\\
 \braket{N|{\mF_{i}}^{\dagger}\mF_{i}|N}&=& g_{i}^{2}(|\beta_{i}|^{2}+N(|\braket{\phi_{B}|\phi_{i}}|^2+|\beta_{i}|^{2}|\braket{\phi_{B}|\psi_{i}}|^{2})).\\ \nonumber
\end{eqnarray}
 If $\ket{\Phi_{B}}=\ket{0}$, we set $N=0$ in the above equations and compute $n_{14 vac}$ from (\ref{ns1}) obtaining the expression
 \beq
 n_{14vac}=g_{1}^{2}g_{2}^{2}t^{4}|\beta_{1}|^{2}(|\braket{\psi_{1}|\phi_{2}}|^{2}-|\beta_{2}|^{2})
 \eeq
 which indicates that the system ${A_{1}A_{2}}$ is entangled for  $0<|\beta_{2}|<|\braket{\psi_{1}|\phi_{2}}|$ ( or equivalently,
$0<|\beta_{1}|<|\braket{\psi_{2}|\phi_{1}}|$). \\We close this section by presenting a situation in which  operators of the form  $\mF_{i}=g_{i}(a(\phi_{i})+\beta_{i}^{*}a^{\dagger}(\psi_{i}))$ appear in the effective interaction between the qubits and the bosonic system.
 Consider two qubits interacting with a relativistic scalar field \cite{Peskin}. Let the Hamiltonian of the system be
  \beq
  \label{hammy}
   \mH=\mH_{0}+\sum_{i=1,2}g_{i}(t)\sigma_{x,i}\hat{\phi}_{i}, \quad  \sigma_{x}=\begin{pmatrix} 0 & 1\\ 1 &0\\ \end{pmatrix}
\eeq
where $\mH_{0}=\frac{1}{2}\sum_{i=1,2}\omega_{A_{i}} \sigma_{z,i}+\sum_{k}\omega_{k}( {a_{k}}^{\dagger}a_{k}+\frac{1}{2})$ is the free Hamiltonian of the qubits+field system. The functions $g_{1}(t)$ and $g_{2}(t)$ describe time dependent coupling strengths and  $\hat{\phi}_{1}=\int_{O_{1}}d^{3}r{}f_{1}(\vec{r})\phi(t,\vec{r})$, $\hat{\phi}_{2}=\int_{O_{2}}d^{3}r{}f_{2}(\vec{r})\phi(t,\vec{r})$ are average fields on the spatial regions $O_{1}$ and $O_{2}$ where the qubits are located.  If the interaction is fast compared to  the free evolution of the qubits and the field (by taking the average fields we introduce a cut-off for the field frequencies) one can make use of Magnus approximation \cite{Magnus} and write the time evolution operator as $\mU=e^{-i\int{dt}\mH_{I}(t)}$ where $\mH_{I}(t)$ is the interaction picture Hamiltonian.
 Following, \cite{Reznik} we assume that the qubits remain causally disconnected  throughout the whole interaction process.
 The time evolution operator can be written as $\mU=e^{-i H_{eff}}$ where
\beq
H_{eff}=\sum_{i=1,2}\int{dt}g_{i}(t)\begin{pmatrix} 0 & e^{-i \omega_{A_{i}}t} \\
                                                    e^{i \omega_{A_{i}}t}&0 \\  \end{pmatrix} \hat{\phi}_{i}(t)=\sum_{i=1,2} \begin{pmatrix} 0 & \mF_{i}^{\dagger} \\ \mF_{i} & 0 \\ \end{pmatrix}.
\eeq

Here the operators $\mF_{i}$ read
\beq
 \mF_{i}=\int_{0}^{\infty}{dt}\, g_{i}(t)e^{i\omega_{A_{i}}t}\hat{\phi}_{i}(t)
=\int\frac{d^3 k}{(2\pi)^{\frac{3}{2}}\sqrt{2\omega_{k}}}\tilde{f_{i}}(k)(\tilde{g}_{i}(\omega_{k}-\omega_{A_{i}})e^{i\vec{k}\cdot \vec{r}_{i}}a_{k}+\gf^{*}_{i}(\omega_{k}+\omega_{A_{i}})e^{-i\vec{k}\cdot \vec{r}_{i}}a_{k}^{\dagger})
\eeq
where $\tilde{g_{i}}(\omega)$ is the Fourier transform of the function $g_{i}(t)$ i.e.  $\tilde{g}_{i}(\omega)=\int{dt}e^{i\omega_{\omega}t}g_{i}(t)$ while $\tilde{f}_{i}(k)$ is the Fourier transform of the smearing function $f_{i}(\vec{r})$. For a massless field and qubits with  $\omega_{A_{1}}=\omega_{A_{2}}$, one can prove that in the limit where the volumes of the regions $O_{1}$ and $O_{2}$ approach zero, one has
\beq
[\mF_{i}, \mF_{i}^{\dagger}]=\int\frac{d^3 k}{(2\pi)^{\frac{3}{2}}\sqrt{2\omega_{k}}}(|\tilde{g}_{i}(\omega_{k}-\omega_{A_{i}})|^{2}-|\tilde{g}_{i}(\omega_{k}+\omega_{A_{i}})|^{2})=
\frac{\omega_{A}}{2\pi}\int{dt} g^{2}(t)\geq 0, \quad i=(1,2).
\eeq
Thus the necessary conditions (\ref{linear}) for the operators $\mF_{i}$  are satisfied. Notice that if we neglect the free evolution of the qubits (we set $\omega_{A_{i}}=0$ ) $\mF_{1}$ and $\mF_{2}$ become normal operators and the state of the qubits remains separable after the interaction. The final entanglement of the two qubit system depends on the form of the functions $g_{i}(t)$, $i=(1,2)$ and it will not be discussed here. See \cite{Reznik} for a detailed discussion.

\section{Entanglement from N particles occupying a single 1-particle state.}
\label{S:N}

\begin{figure}[htb]
\centering
\subfigure[ N=1 (dashed line) and N=3 (solid line)]{\includegraphics[scale=1]{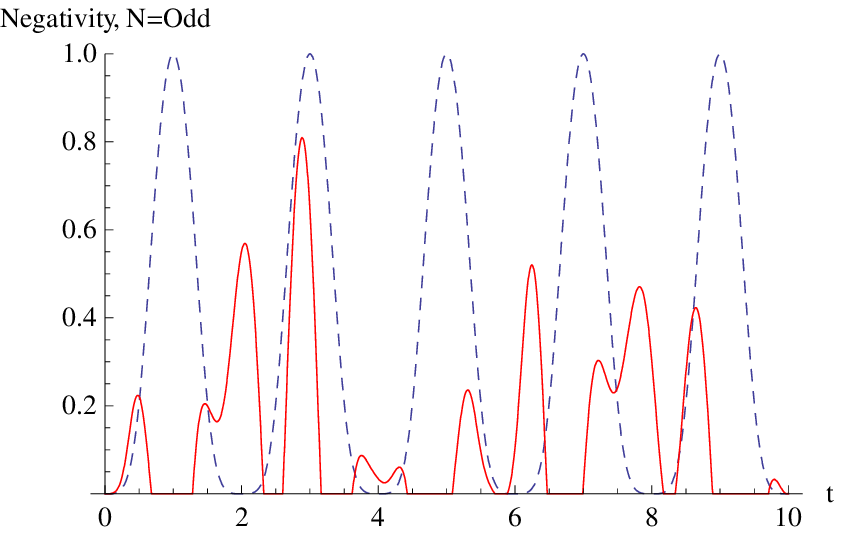}{\label{F:Nodd}}}
\subfigure[ N=2 (dashed line) and N=4 (solid line)]{\includegraphics[scale=1]{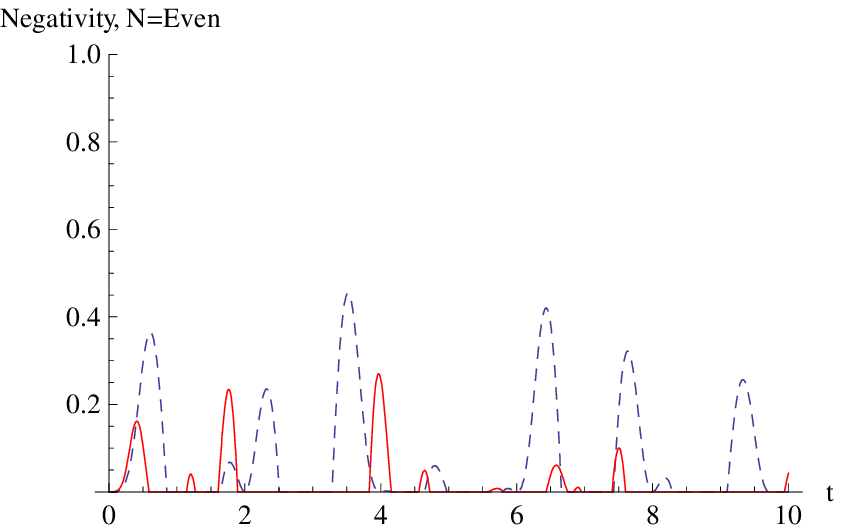}\label{F:N}}
\caption{Negativity as a function of time for odd and even values of N. We have assumed $g_{1}=g_{2}=\frac{\pi}{2}$ and $u_{1}=u_{2}=\frac{1}{\sqrt{2}}.$}
\end{figure}

In the previous section we found that for interactions of the form $(\ref{special})$ with $\mF_{i}=g_{i}a(\phi_{i})$,  the system $A_{1}A_{2}$, initially in the state, $\ket{a_{1}}=\ket{0,0}$,  becomes instantaneously entangled when B contains N particles occupying a single 1-particle state $\ket{\phi_{B}}$. The only requirement for the state $\ket{\phi_{B}}$ is to overlap with $\ket{\phi_{1}}$ and $\ket{\phi_{2}}$ i.e. $u_{1}=\braket{\phi_{1}|\phi_{B}}\neq 0$ and $u_{2}=\braket{\phi_{2}|\phi_{B}}\neq 0$. Since the states  $\ket{\phi_{1}}$ and $\ket{\phi_{2}}$ are orthogonal, any 1-particle state can be written as $\ket{\phi_{B}}=\sum_{i} u_{i}a_{i}^{\dagger}\ket{0}+u_{T}\ket{\phi_{T}}$ where $\braket{\phi_{i}|\phi_{T}}=0$ for $i=(1,2).$ Therefore
 the state $\ket{\Phi_{B}}=\frac{1}{{\sqrt{N!}}}{a^{\dagger}(\phi_{B})}^{N}\ket{0}$ can be written in occupation number representation as
\beq
\label{occupation}
\ket{\Phi_{B}}=\sum_{n_{1},n_{2}}\sqrt{\frac{N!}{n_{1}! n_{2}! (N-n_{1}-n_{2})!}}u_{1}^{n_{1}} u_{2}^{n_{2}}u_{T}^{N-n_{1}-n_{2}}\ket{n_{1},n_{2},N-n_{1}-n_{2}}
\eeq
Now making use of (\ref{series1}), (\ref{series2}) and (\ref{bigmatrix}), we compute the  reduced density matrix for  $A_{1}A_{2}$ as a function of time.  For example, the diagonal element
\beq
\rho_{11}=\braket{\mK_{1}^{\dagger}\mK_{1}{\mK_{2}}^{\dagger}{\mK_{2}}}=\braket{\Phi_{B}|
\cos^{2}(\sqrt{{\mF_{1}}^{\dagger}\mF_{1}}t)\cos^{2}(\sqrt{{\mF_{2}}^{\dagger}\mF_{2}}t)|\Phi_{B}}
\eeq
can be easily computed using the occupation number representation  $(\ref{occupation})$ for the state $\ket{\Phi_{B}}$. Likewise, one may determine the remaining  diagonal entries of $\rho^{A}$  which can written in the compact form
\beq
\label{Nkk}
\rho_{kk}=\sum_{n_{1},n_{2}=0}^{N}\frac{N!}
{n_{1}!n_{2}! (N-n_{1}-n_{2})!}P_{n_{1},n_{2}}(u_{1},u_{2})F_{kk}(n_{1},n_{2})
\eeq
where $P_{n_{1}, n_{2}}(u_{1},u_{2})=|u_{1}|^{2n_{1}}|u_{2}|^{2n_{2}}(1-|u_{1}|^{2}-|u_{2}|^{2})^{N-n_{1}-n_{2}}$ and \begin{eqnarray}
F_{11}(n_{1},n_{2})&=&\cos^{2}(\sqrt{n_{1}}g_{1}t)\cos^{2}(\sqrt{n_{2}}g_{2}t), \quad F_{33}(n_{1},n_{2})=\sin^{2}(\sqrt{n_{1}}g_{1}t)\cos^{2}(\sqrt{n_{2}}g_{2}t), \\
F_{22}(n_{1},n_{2})&=&\cos^{2}(\sqrt{n_{1}}g_{1}t)\sin^{2}(\sqrt{n_{2}}g_{2}t), \quad F_{44}(n_{1},n_{2})=\sin^{2}(\sqrt{n_{1}}g_{1}t)\sin^{2}(\sqrt{n_{2}}g_{2}t). \\ \nonumber
\end{eqnarray}
 From equation (\ref{bigmatrix}), it can be easily seen that operators of the form $\mF_{i}=g_{i}a_{i}$ lead to $\rho_{14}=0$.  Hence, the only non-vanishing off-diagonal matrix elements are $\rho_{23}=\braket{\Phi_{B}|{\mN_{1}}^{\dagger}\mK_{1}{\mK_{2}}^{\dagger}\mN_{2}|\Phi_{B}}$
and, obviously, $\rho_{32}={\rho_{23}}^{*}.$ Again, using (\ref{occupation}), one obtains
\beq
\label{N23}
\rho_{23}=u_{1}^{*}u_{2}\sum_{n_{1},n_{2}=0}^{N-1}\frac{N!}{n_{1}! n_{2}! (N-n_{1}-n_{2}-1)!} Q_{n_{1},n_{2}}(u_{1},u_{2})F_{23}(n_{1},n_{2})
\eeq
where the polynomial $Q_{n_{1},n_{2}}(u_{1},u_{2})=|u_{1}|^{2n_{1}}|u_{2}|^{2n_{2}}(1-|u_{1}|^2-|u_{2}|^2)^{N-n_{1}-n_{2}-1}$ and

\beq
F_{23}(n_{1},n_{2})=\cos(\sqrt{n_{1}}g_{1}t)\frac{\sin(\sqrt{n_{1}+1}g_{1}t)}{\sqrt{n_{1}+1}}\cos(\sqrt{n_{2}}g_{2}t)\frac{\sin(\sqrt{n_{2}+1}g_{2}t)}{\sqrt{n_{2}+1}}.\eeq
 Unfortunately, it does not seem possible to express the sums $(\ref{Nkk})$ and $(\ref{N23})$ in closed form. However, numerical analysis indicates that one obtains entangled states for the qubits for all values of $N$ with larger values of negativity when N is an odd number (see Fig.{\ref{F:N}}). The above results can be extended to the case where
\beq
\mF_{i}=g_{i}a_{i}^{m}, \quad i=(1,2),\quad m\geq 1,  \quad  \textrm{and} \quad   [a_{1},{a_{2}}^{\dagger}]=0.
 \eeq
 This type of operators describe the situation where $m$ particles  are needed to flip (or excite one of the atoms) the state of one of the qubits  \cite{Singh}. The expressions $({\ref{Nkk}})$ and $({\ref{N23}})$ (corresponding to the case m=1) can be generalized to any value of m. Using the identities $a_{i}{{a_{i}}^{\dagger}}^{m} =\frac{({a_{i}}^{\dagger}a_{i}+m)!}{{(a_{i}}^{\dagger}a_{i})!}$, $ {a_{i}}^{\dagger m}a_{i}=\frac{({a_{i}}^{\dagger}a_{i})!}{({a_{i}}^{\dagger}a_{i}-m)!}$ and writing the state $\ket{\Phi_{B}}$ as in equation (\ref{occupation}), one arrives at the following expressions:
\begin{eqnarray}
\label{Nmkk}
\rho_{kk}(m)&=&\sum_{n_{1},n_{2}}\frac{N!}{n_{1}!n_{2}!(N-n_{1}-n_{2})!}{P}_{n_{1},n_{2}}(u_{1},u_{2}){\stackrel{(m)}{{F}}}_{kk}(n_{1},n_{2}), \quad  \\
\label{Nm23}
\rho_{23}(m)&=&(u_{1}^{*}u_{2})^{m}\sum_{n_{1},n_{2}}\frac{N!}{n_{1}!n_{2}!(N-n_{1}-n_{2}-m)!}{\stackrel{(m)}{Q}}_{n_{1},n_{2}}(u_{1},u_{2}){\stackrel{(m)}{F}}_{23}(n_{1},n_{2})\\ \nonumber \end{eqnarray}
where   $\stackrel{(m)}{F_{kk}}(n_{1},n_{2})=F_{kk}({\frac{n_{1}!}{(n_{1}-m)!}},{\frac{n_{2}!}{(n_{2}-m)!}})$
, ${\stackrel{(m)}{Q}}_{n_{1},n_{2}}(u_{1},u_{2})=|u_{1}|^{2n_{1}}|u_{2}|^{2n_{2}}(1-|u_{1}|^2-|u_{2}|^2)^{N-n_{1}-n_{2}-m}$ and

\beq
\stackrel{(m)}{F_{23}}(n_{1},n_{2})=\frac{\cos(\sqrt{\frac{n_{1}!}{(n_{1}-m)!}}g_{1}t)\sin(\sqrt{\frac{(n_{1}+m)!}{n_{1}!}}g_{1}t)}
{\sqrt{\frac{(n_{1}+m)!}{n_{1}!}}}\frac{\cos(\sqrt{\frac{n_{2}!}{(n_{2}-m)!}}g_{2}t)\sin(\sqrt{\frac{(n_{2}+m)!}{n_{2}!}}g_{2}t)}
{\sqrt{\frac{(n_{2}+m)!}{n_{2}!}}}.
\eeq

From equation $(\ref{Nmkk})$ one finds that  $\rho_{44}$ vanishes for states having $N=\{m,m+1, \ldots, 2m-1\}$ particles occupying the 1-particle state $\ket{\phi_{B}}$. This is easy to understand; the matrix element $\rho_{44}$ corresponds to the process in which the states of both qubits are flipped. Therefore, the presence of at least $2m$ particles is required to have $\rho_{44}\neq 0$. In our model,  $\rho_{44}(t)=0$ implies that the two qubits are entangled with negativity $\mathcal{N}(\rho^{A})=\sqrt{(\rho_{11})^2+ 4|\rho_{23}|^2}-\rho_{11} \geq 0$ for any value of t (except for those values of t for which $\rho_{23}(t)=0$).
 For $N=m$, $|u_{1}|=|u_{2}|=\frac{1}{\sqrt{2}}$ and  $g_{1}=g_{2}$, expressions (\ref{Nmkk}) and (\ref{Nm23}) take the particulary simple form
\beq
\rho_{23}=\frac{1}{2^{N}}\sin^{2}(\sqrt{N!}gt), \quad \rho_{kk}=\frac{N!}{2^{N}}\sum_{n_{1}+n_{2}=N}\frac{{\stackrel{(m)}{F}_{kk}}(n_{1},n_{2})}{n_{1}!n_{2}!} \quad \textrm{for} \quad k=(1,2,3) \quad \textrm{and} \quad \rho_{44}=0.
\eeq
 Here, we notice that the maximum values of the negativity $\mathcal{N}(\rho^{A})$, behave like $1/{2^{2N-1}}$ for large values of N. For $m>1$, this behavior can be improved. In fact, for $N=m+1$ we have $\rho_{23}=\frac{\sqrt{N}}{2^{N-1}}\sin(\sqrt{m!}gt)\sin(\sqrt{(m+1)!}gt)$, $\rho_{44}=0$ and as a result  $\mathcal{N(\rho^{A})} \sim N/2^{{(2N-3)}}$ for large values of N.
 Graphs for the cases $m=2$, $N=2$ and  $m=2$, $N=3$ are shown in Fig.$\ref{F:m=2}$. In what follows, we will restrict our discussion to the case $m=1$.

\subsection{Entanglement from 1-Particle States}
\begin{figure}[h]
\centering
\subfigure[ Cases $N=2$, $m=2$ (solid line) and $N=3$,$m=2$ (dashed line). Here $|u_{1}|=|u_{2}|=\frac{1}{\sqrt{2}}$ and $g_{1}=g_{2}=1$. ]{\includegraphics{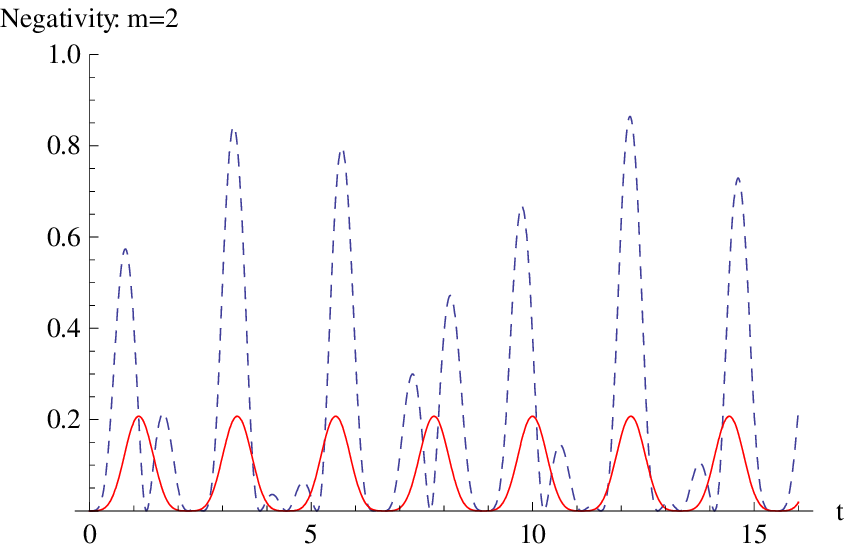} {\label{F:m=2}}}
\subfigure[Negativity for $\ket{\phi_{A}}=\ket{1,1}$. Here $|u_{1}|=|u_{2}|=\frac{1}{\sqrt{2}}$ and $g_{1}=g_{2}=\frac{\pi}{2}$. The maximum value found is   $\mathcal{N}=0.98. $]{\includegraphics{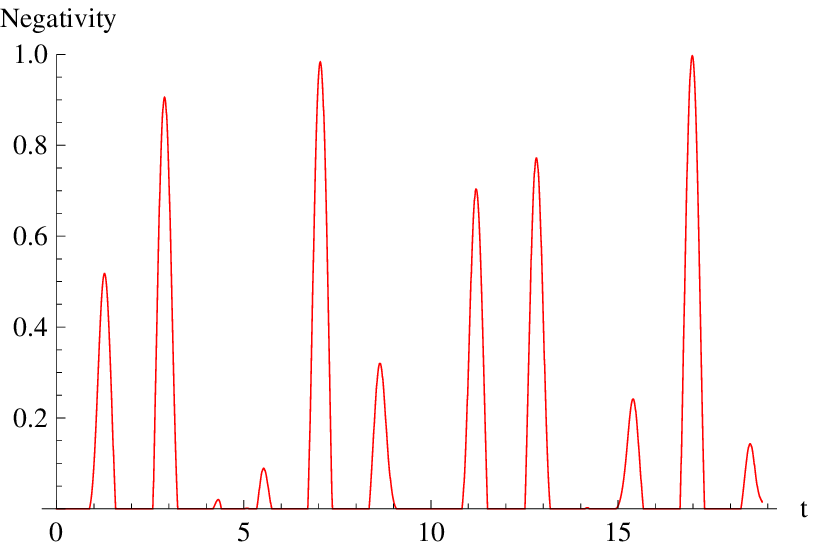}{\label{F:11}}}
\caption{Negativity vs. time.}
\end{figure}

  If the qubit system is initially in the state $\ket{0,0}$, we obtain from $(\ref{Nkk})$ and $(\ref{N23})$ the following density matrix
 \begin{align*}
\rho_{11}&=1-|u_{1}|^2\sin^{2}(g_{1}t)- |u_{2}|^2\sin^{2}(g_{2}t)&
\rho_{22}=|u_{2}|^{2}\sin^{2}(g_{2}t)\\
\rho_{33}&=|u_{1}|^{2}\sin^{2}(g_{1}t) & \rho_{23}={u_{1}}^{*}u_{2}\sin(g_{1}t)\sin(g_{2}t). \\
\end{align*}
 As we know from previous discussions, this state is always entangled with negativity $\mathcal{N}(\rho^{A})=\sqrt{(\rho_{11})^2+ 4|\rho_{23}|^2}-\rho_{11}$. Furthermore, if the amplitudes $u_{i}$ satisfy
 $|u_{1}|=|u_{2}|=\frac{1}{\sqrt{2}}$ and $g_{1}=g_{2}$, the negativity oscillates between zero and its maximum value $\mathcal{N}=1$ with period $\frac{\pi}{g}$ (see Fig.\ref{F:Nodd}). However, this behavior changes substantially
  if the initial state of $A_{1}A_{2}$ is  $\ket{a_{4}}=\ket{1,1}$. From section (\ref{S:Operators}), we know that the reduced density matrix can be obtained from the relation
  \begin{equation}
  {\rho^{A}}(\ket{a_{4}}, \mU) =\begin{pmatrix}
   \tilde{\rho}_{44}&0&0&\tilde{{\rho}}_{41}\\
   0&\tilde{{\rho}}_{33}&\tilde{{\rho}}_{32}&0\\
   0&\tilde{\rho}_{32}^{*}&\tilde{{\rho}}_{22}&0\\
   \tilde{\rho}_{14}&0&0&\tilde{\rho}_{11}\\
   \end{pmatrix}=V\rho^{A}(\ket{a_{1}},\mU({\mF_{1}}^{\dagger},{\mF_{2}}^{\dagger})V^{\dagger}, \quad V=\sigma_{x}\otimes \sigma_{x}.
   \end{equation}
   In this case we also have ${\tilde{\rho}}_{14}=0$ and therefore in order to quantify the entanglement in system $A_{1}A_{2}$ one needs the following matrix elements:
\begin{eqnarray}
\label{111}
\rho_{44}&=&\cos^2(g_{1}t)\cos^2(g_{2}t)+|u_{1}|^2 \cos^2(g_{2}t)(\cos^2(\sqrt{2}g_{1}t)-\cos^2(g_{1}t))\\ \nonumber
&+&|u_{2}|^2 \cos^2(g_{1}t)(\cos^2(\sqrt{2}g_{2}t)-\cos^2(g_{2}t))
\\
\label{112}
\rho_{11}&=&\sin^{2}(g_{1}t)\sin^{2}(g_{2}t)-|u_{1}|^2 \sin^2(g_{2}t)(\cos^2(\sqrt{2}g_{1}t)-\cos^2(g_{1}t))\\
\nonumber
&-&|u_{2}|^2 \sin^2(g_{1}t)(\cos^2(\sqrt{2}g_{2}t)-\cos^2(g_{2}t))\\
\label{113}
\rho_{32}&=&u_{1}u^{*}_{2}\cos(\sqrt{2}g_{2}t)\cos(\sqrt{2}g_{1}t)\sin(g_{1}t)\sin(g_{2}t).\\ \nonumber
\end{eqnarray}
  The above expressions dictate the time dependence of the negativity (see Fig.{\ref{F:11}}) and contrary to the situation in which the initial state of $A_{1}A_{2}$ was $\ket{0,0}$, now system $A_{1}A_{2}$ exhibits periods of entanglement death and entanglement revivals. Nevertheless, for the symmetric scenario ( $g_{1}=g_{2}=g$ and $|u_{1}|=|u_{2}|=\frac{1}{\sqrt{2}}$) one can obtain an almost maximally entangled state ($\mathcal{N}\approx 1$). Notice from (\ref{111}, \ref{112}, \ref{113}) that one can simultaneously have $\rho_{11} \approx 0$, $\rho_{44}\approx 0$ and $ \rho_{23} \approx \frac{1}{2}$ when
   \begin{equation}
  \sqrt{2}gt \approx n \pi \quad  \textrm{and} \quad  gt \approx(2m+1)\frac{\pi}{2} \Longrightarrow (2m+1)\approx n\sqrt{2}.
   \end{equation}
 The first three pairs  of numbers of the form $(n, 2m+1)$ satisfying (approximately) these equations are (5,7), (12,17) and (29, 41), ( $5\sqrt{2}=7.07, \quad12 \sqrt{2}=16.97,\quad 29 \sqrt{2}=41.01$).  The first two pairs correspond to the peaks with $\mathcal{N}\approx 1 $ in Fig.\ref{F:11}. At these points the system $A_{1}A_{2}$ is in the state $\rho^{A}\approx \ket{\psi^{+}}\bra{\psi^{+}}$ with $\ket{\psi^{+}}=\frac{1}{2}(\ket{a_{2}}+\ket{a_{3}})$.

\subsection{Operators of the form $\mF_{i}=g_{i}(a_{i}+|\beta_{i}|e^{i\theta_{i}}{a_{i}}^{\dagger})$}
 Another type of operators $\mF_{i}$ belonging to the class $(\ref{class})$ are the operators of the form
  $\mF_{i}=g_{i}(a_{i}+|\beta_{i}|e^{i\theta_{i}}{a_{i}}^{\dagger})$ with $|\beta_{i}|<1$ and $[a_{1},{a_{2}}^{\dagger}]=0$.  For simplicity,  we assume that system $A_{1}A_{2}$ is initially in the state $\ket{0,0}$.  For these operators, one can sum the series $(\ref{series1})$ and $(\ref{series2})$ by means of a Bogoliubov transformation \cite{Fetter}. In fact, one may set $\beta_{i}=\tanh(r_{i})$ and write  $\mF_{i}={\tilde{g}_{i}}b_{i}$ where $b_{i}=\cosh(r_{i})a_{i}+e^{i\theta_{i}}\sinh(r_{i}){a_{i}}^{\dagger}$ and $\tilde{g}_{i}=\frac{g_{i}}{\cosh(r_{i})}$. Using the identities
 \begin{equation}
 b_{i}=\mathcal{U}_{i}a_{i}\mathcal{U}_{i}^{\dagger}, \quad \textrm{with}\quad \mathcal{U}_{i}=e^{\frac{1}{2}({z_{i}}^{*}{a_{i}}^2-{z_{i}} {{a_{i}}^{\dagger}}^{2})},\quad  z=r_{i}e^{i \theta_{i}}
 \end{equation}
one can express the vacuum corresponding to the operators $a_{i}$ in terms of the eigenstates of the operators $n_{i}={b_{i}}^{\dagger}b_{i}$. That is
\begin{equation}
\ket{\bar{0}}=\sum_{n_{1},n_{2}}c_{n_{1},n_{2}}\ket{2n_{1},2n_{2},0}, \quad c_{n_{i}}=\frac{1}{\sqrt{\cosh(r_{i})}}e^{i n_{i} \theta_{i}}\tanh^{n_{i}}(r_{i})\frac{\sqrt{(2n_{i})!}}{2^{n_{i}}n_{i}!}.
\end{equation}
In this new basis the 1-particle excitation $\ket{\Phi_{B}}=a_{B}^{\dagger}\ket{0}$ with $u_{1}=u_{2}=\frac{1}{\sqrt{2}}$ assumes the form
\begin{equation}
\ket{\Phi_{B}}=\frac{1}{\sqrt{2}}\sum_{n_{1},n_{2}}c_{n_{1}}c_{n_{2}}(\frac{\sqrt{2n_{1}+1}}{\cosh(r_{1})}\ket{2n_{1}+1, 2n_{2}}
+\frac{\sqrt{2n_{2}+1}}{\cosh(r_{2})}\ket{2n_{1},2n_{2}+1}).
\end{equation}
 Assuming $g_{1}=g_{2}$ and taking into account the fact that the states $\ket{n_{i}}$ are eigenstates of the operators ${\mF_{i}}^{\dagger}\mF_{i}$ with eigenvalues ${{\tilde{g}}_{i}}^{2} n_{i}$ we obtain:
\begin{equation}
\rho^{A}=\left(
  \begin{array}{cccc}
    AB & 0 & 0 & -EF  \\
    0 & \frac{AD+BC}{2} & \frac{E^{2}+F^{2}}{2} & 0\\
    0 & \frac{E^{2}+F^{2}}{2} &\frac{AD+BC}{2} & 0 \\
    -EF & 0 & 0 & CD \\
  \end{array}
\right)
\end{equation}
with A,B,C,D,E,F given by the following series
\begin{eqnarray}
A=\sum_{n} \frac{2n+1}{\cosh^{2}(r)}|c_{n}|^2\cos^{2}(\sqrt{2n+1}\tilde{g}t), \quad B=\sum_{n} |c_{n}|^{2}\cos^{2}(\sqrt{2n}\tilde{g}t)\\
C=\sum_{n}\frac{2n+1}{\cosh^{2}(r)}|c_{n}|^2\sin^{2}(\sqrt{2n+1}\tilde{g}t), \quad D=\sum_{n}|c_{n}|^2\sin^2(\sqrt{2n}\tilde{g}t)\\\nonumber
\end{eqnarray}
and
\begin{eqnarray}
E&=&\sum_{n}\frac{\sqrt{2n+1}}{\cosh(r)}|c_{n}|^2\cos(\sqrt{2n}\tilde{g}t)\sin(\sqrt{2n+1}\tilde{g}t)\\
F&=&\frac{\sinh(r)}{\cosh(r)^2}\sum_{n}\frac{2n+1}{\sqrt{2(n+1)}}|c_{n}|^2\cos(\sqrt{2n+1}\tilde{g}t)\sin(\sqrt{2(n+1)}\tilde{g}t)
\\ \nonumber
\end{eqnarray}

\begin{figure}[htb]
\centering
\includegraphics[scale=1]{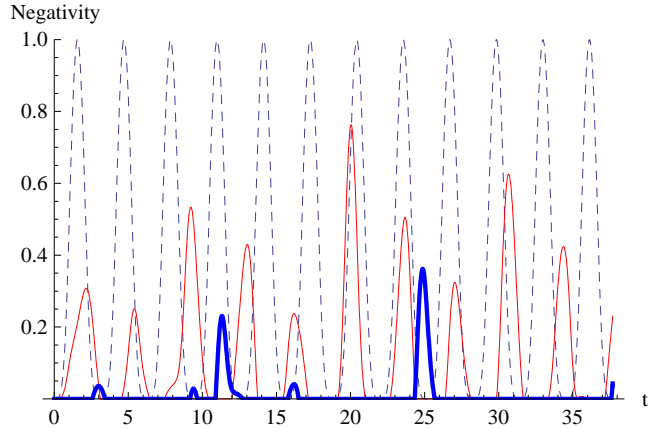}
\caption{Negativity vs. time for  different values of
$\beta$. Here we assume $g_{1}=g_{2}=1$ and $u_{1}=u_{2}=\frac{1}{\sqrt{2}}$. The cases $\beta=0$, $\beta=0.5$, and $\beta=0.7$ are represented by dashed, solid, and thick solid lines respectively.}
\label{F:bogo}
\end{figure}

Notice, that in the limit $|\beta|=1$, one has $[\mF_{i},{\mF_{i}}^{\dagger}]=0$. Hence, according to section (\ref{S:Operators}), the  entanglement in system A should disappear as we approach $|\beta|=1$. In Fig.{\ref{F:bogo}}
we present graphs of entanglement versus time for different values of $|\beta|.$ Numerical analysis shows, that entanglement is more strongly deteriorated for values of $\beta$ close to 1. Thus we conclude that operators being mixtures of the form $a+{\beta}a^{\dagger}$ with $\beta<1$ can also transfer a substantial amount of entanglement to the two qubit system.

\subsection{Entanglement from Mixed States}
The expressions $(\ref{Nkk})$ and $(\ref{N23})$ can be also used to determine the final state of the qubits when system B is in a mixed state of the form $\rho_{B}=\sum_{N}p_{N}\ket{N}\bra{N}$. In this case, the matrix elements of $\rho^{A}$ read
 \beq
 \label{PN}
  \rho^{A}_{ij}=\sum_{N}p_{N}\rho_{ij}(N)
  \eeq
with $\rho_{ij}(N)$  given by equations $(\ref{Nkk})$ and $(\ref{N23})$. The separability of the qubits $A_{1}A_{2}$ depends on the distribution $\{p_{N}\}$. For example, consider the binomial distribution $p_{N}=\binom{M}{N}p^{N}(1-p)^{M-N}$ (which was already discussed in section(\ref{S:Expansion})). In this case, equations (\ref{PN}), $(\ref{Nkk})$ and $(\ref{N23})$ yield
\begin{equation}
\rho^{A}_{kk}=\sum_{n_{1},n_{2}}\frac{M!}{n_{1}! n_{2}!(M-n_{1}-n_{2})!}(\sqrt{p}|u_{1}|)^{2n_{1}}(\sqrt{p}|u_{2}|)^{2 n_{2}}(1-(\sqrt{p}|u_{1}|)^{2}-(\sqrt{p}|u_{2}|)^{2})^{M-n_{1}-n_{2}}F_{k,k}(n_{1},n_{2}).
 \end{equation}
  Likewise, we compute $\rho^{A}_{23}$ from $(\ref{N23})$   to find that the matrix elements $\rho^{A}_{i,j}$ corresponding to the binomial distribution $p_{N}=\binom{M}{N}p^{N}(1-p)^{M-N}$ have the same form as the $\rho_{ij}(N)$ from (\ref{Nkk}) and (\ref{N23}) with N replaced by M and the amplitudes $u_{i}$ replaced by $\sqrt{p}u_{i}.$ Therefore, this case reduces to the previously studied situation where  $\rho_{B}=\ket{M}\bra{M}$. On the other hand, since entanglement disappears as the amplitudes $u_{i}=\braket{\phi_{i}|\phi_{B}}$ approach zero, one expect to obtain a separable state $\rho^{A}$ for a Poissonian distribution i.e. $\rho_{B}= p_{N}\ket{N}\bra{N}$ with $p_{N}=\frac{\lambda^{N}}{N!}e^{-\lambda}$ (recall that Poisson distribution is the limit of the binomial distribution for $M\rightarrow \infty$, $p\rightarrow 0$ and $M p \rightarrow \lambda$). In fact, for a Poissonian distribution, one obtains from $(\ref{PN})$  $(\ref{Nkk})$ and $(\ref{N23})$ the state
\begin{equation}
\label{Poisson}
\rho^{A}=\begin{pmatrix} c_{1}c_{2}&0&0&0\\
                         0 & c_{1}s_{2}& {m_{1}}^{*}m_{2}&0\\
                         0& m_{1}{m_{2}}^{*}& c_{2}s_{1}&0 \\
                         0 & 0 &0&s_{1}s_{2}\\
                         \end{pmatrix}
                         \end{equation}
where the functions $c_{i}$, $s_{i}$, and $m_{i}$  given by
\begin{eqnarray*}
c_{i}&=&e^{-\lambda |u_{i}|^{2}}\sum_{n_{i}}\frac{(\lambda|u_{1}|^2)^{n_{i}}}{n_{i}!}\cos^{2}(\sqrt{n_{i}}g_{i}t),  \quad   s_{i}=e^{-\lambda |u_{i}|^{2}}\sum_{n_{i}}\frac{(\lambda|u_{i}|^2)^{n_{i}}}{n_{i}!}\sin^{2}(\sqrt{n_{i}}g_{i}t),\\
m_{i}&=&\sqrt{\lambda}u_{i}e^{-\lambda |u_{i}|^{2}}\sum_{n_{i}}\frac{(\lambda|u_{1}|^2)^{n_{i}}}{n_{i}!}\cos(\sqrt{n_{i}}g_{i}t)\frac{\sin(\sqrt{n_{i}+1}g_{i}t)}{\sqrt{n_{i}+1}}.
\end{eqnarray*}

A matrix with the structure of (\ref{Poisson}) must necessarily be separable. Notice that the entanglement condition $|\rho_{23}|^{2}>\rho_{11}\rho_{44}$ is not compatible with the positivity condition $|\rho_{23}|^{2}<\rho_{22}\rho_{33}$. Hence, $\rho^{A}$ is separable.

\section{Entanglement from N particles occupying different 1-particle states}
\label{S:Ndiff}

So far we have considered N-particle excitations of system $B$ with all the particles occupying the same 1-particle state. It is also interesting to study  multiparticle states of the form
 \begin{equation}
 \label{ortho}
\ket{\Phi_{B}}=\prod_{k=1}^{N} a^{\dagger}({\phi_{B}}_{k})\ket{0}, \quad  \braket{{\phi_{B}}_{k}|{\phi_{B}}_{k'}}=\delta_{k,k'}
\end{equation}
representing N identical particles occupying mutually orthogonal 1-particle states. Again, we assume that the interactions between B and $A_{i}$ are of  the form (\ref{special}) with  $\mF_{i}=g_{i}a(\phi_{i})$ and $\braket{\phi_{1}|\phi_{2}}=0$. It is clear that now the entanglement transferred to $A_{1}A_{2}$
depends on the relative geometry of the set of states $\{\phi_{B_{1}},\phi_{B_{2}}, \ldots, {\phi_{B_{N}}}\}$ and the states $\ket{\phi_{1}}$ and $\ket{{\phi_{2}}}$. One can find the density matrix for states of the form (\ref{ortho}) using expressions (\ref{Nkk}) and (\ref{N23}). Taking the linear combinations $\ket{\phi_{B}}=\sum_{k=1}^{N} x_{k}\ket{\phi_{B_{k}}}$, $\ket{\tilde{\phi}_{B}}=\sum_{k=1}^{N} y_{k}^{*}\ket{\phi_{B_{k}}}$ and defining the states $\ket{\Phi_{B}}\equiv \frac{1}{\sqrt{N!}}{a^{\dagger}}^{N}(\phi_{B})\ket{0}$, $\ket{\tilde{\Phi}_{B}}\equiv \frac{1}{\sqrt{N!}}{a^{\dagger}}^{N}(\tilde{\phi}_{B})\ket{0}$, one computes the auxiliary matrix element
\beq
\tilde{\rho}_{11}(x,y)\equiv\braket{\tilde{\Phi}_{B}|{\mK_{1}}^{\dagger}\mK_{1}\mK_{2}^{\dagger}\mK_{2}|\Phi_{B}}=\sum_{n_{1},n_{2}}\frac{N!}{n_{1}!n_{2}!(N-n_{1}-n_{2})!}P_{n_{1},n_{2}}(x,y)F_{11}(n_{1},n_{2}).
\eeq
 In the above expression, the polynomial $P_{n_{1},n_{2}}(x,y)$ is given by
\beq
P_{n_{1},n_{2}}(x,y)=(u_{1}\tilde{u}_{1}^{*})^{n_{1}}(u_{2} {\tilde{u}_{2}}^{*})^{n_{2}}(\braket{\tilde{\phi}_{B}|\phi_{B}}-u_{1}\tilde{u}_{1}^{*}-u_{2} {\tilde{u}_{2}}^{*})^{N-n_{1}-n_{2}}.
\eeq
with $u_{i}=\sum_{k} x_{k}u_{i,k}$, $\tilde{u}_{i}^{*}=\sum_{k}y_{k}u_{i,k}^{*}$ and $u_{i,k}\equiv\braket{\phi_{i}|\phi_{B_{k}}}$.
 From the above expression we can extract the first diagonal element of the two qubit density matrix. Notice, that  $\rho_{11}$  (which corresponds to the original state $\ket{\Phi_{B}}=\prod_{k=1}^{N} a^{\dagger}({\phi_{B}}_{k})\ket{0}$)  is related to $\tilde{\rho}_{11}(x,y)$ as follows:
\begin{eqnarray}
\label{r11}
\rho_{11}&=&\frac{1}{N!}\frac{\partial^{2N}}{\partial{x_{1}}\ldots \partial{x_{N}}\partial{y_{1}}\ldots \partial{y_{N}}}\tilde{\rho}_{1,1}(x,y)\nonumber \\
&=&\sum_{n_{1},n_{2}}\frac{1}{n_{1}!n_{2}!(N-n_{1}-n_{2})!}F_{11}(n_{1},n_{2})\frac{\partial^{2N}}{\partial{x_{1}}\ldots \partial{x_{N}}\partial{y_{1}}\ldots \partial{y_{N}}}P_{n_{1},n_{2}}(x,y).
\end{eqnarray}
Following the same steps, one obtains $\rho_{22}, \rho_{33}$ and $\rho_{44}$. Similarly, one finds that the off diagonal element $\rho_{23}$ is
\begin{eqnarray}
\label{r23}
{\rho}_{23}&=&\frac{1}{N!}\frac{\partial^{2 N}}{\partial{x_{1}}\ldots \partial{x_{n}}\partial{y_{1}}\ldots \partial{y_{N}}}\tilde{\rho}_{23}(x,y)\nonumber \\
&=&\sum_{n_{1},n_{2}}\frac{1}{n_{1}! n_{2}!(N-n_{1}-n_{2}-1)! } F_{23}(n_{1},n_{2}) \frac{\partial^{2 N}}{\partial{x_{1}}\ldots \partial{x_{n}}\partial{y_{1}}\ldots \partial{y_{N}}}Q_{n_{1},n_{2}}(x,y)\nonumber \\
\end{eqnarray}
with the polynomial  $Q_{n_{1},n_{2}}(x,y)$ given by
  \beq
 Q_{n_{1},n_{2}}(x,y)={\tilde{u}_{1}}^{*}u_{2}(u_{1}\tilde{u}_{1}^{*})^{n_{1}}(u_{2} {\tilde{u}_{2}}^{*})^{n_{2}}(\braket{\tilde{\phi}_{B}|\phi_{B}}-u_{1}\tilde{u}_{1}^{*}-u_{2} {\tilde{u}_{2}}^{*})^{N-n_{1}-n_{2}-1}.
 \eeq
 Using the above equations one can express the density matrix $\rho^{A}$ in terms of  the $2\times N$ matrix $u_{i,k}=\braket{\phi_{i}|\phi_{B_{k}}}$ and the constants $g_{1}$ and $g_{2}$. Let us study the particular case where $\ket{\phi_{B_{1}}}$ and $\ket{\phi_{B_{2}}}$ lie on the plane spanned by $\ket{\phi_{1}}$ and $\ket{\phi_{2}}$. Then they are related by an $SU(2)$ transformation, i.e.

 \begin{equation}
 \label{su(2)}
  \begin{pmatrix} \phi_{B_{1}}\\
                                                \phi_{B_{2}}\\
\end{pmatrix}=\begin{pmatrix} \cos(\theta)& \sin(\theta)e^{i\eta}  \\
                              -\sin(\theta)e^{-i\eta} & \cos(\theta)\\ \end{pmatrix}\begin{pmatrix}
                                                                                 \phi_{1}\\ \phi_{2}\end{pmatrix}.
\end{equation}
Making use of (\ref{r23}) one obtains
\begin{equation}
 \rho_{23}(t)=\frac{1}{2\sqrt{2}}\sin(4\theta)e^{i\eta}(\sin(g_{1}t)\sin(\sqrt{2}g_{2}t)\cos(g_{2}t)-\sin(g_{2}t)
 \sin(\sqrt{2}g_{1}t)\cos(g_{1}t))
 \end{equation}
 which vanishes when either $\sin(4 \theta)=0$ or  $g_{1}=g_{2}.$ Therefore, in this case, the entanglement transfer scheme works if the coupling constants are different. The diagonal elements $\rho_{11}$ and $\rho_{44}$ read
\begin{eqnarray}
\rho_{11}(t)&=&\frac{1}{2}\sin^2(2\theta)(\cos^2(\sqrt{2}g_{1}t)+\cos^2(\sqrt{2}g_{2}t))+\cos^2(2\theta)\cos^{2}(g_{1}t)
\cos^2(g_{2}t)\\
\rho_{44}(t)&=&\cos^{2}(2 \theta)\sin^{2}(g_{1}t)\sin^2(g_{2}t).\\ \nonumber
\end{eqnarray}

\begin{figure}[htb]
\centering
\includegraphics[scale=0.9]{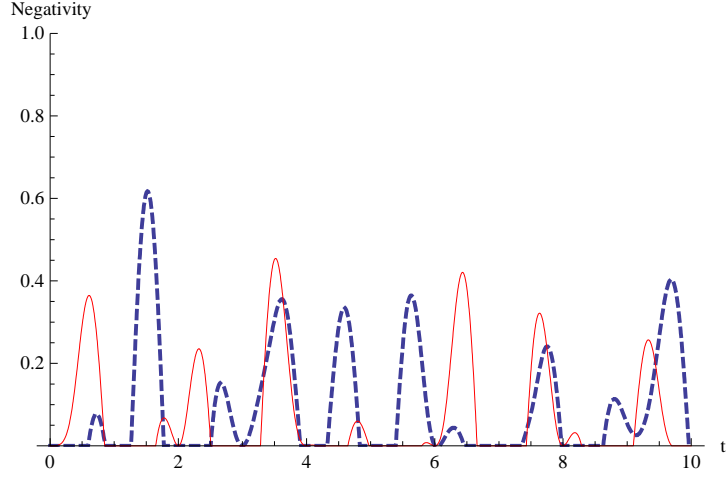}
\caption{Negativity as a function of time. The dashed line corresponds to a state of the form (\ref{su(2)}) with  $\theta=\frac{\pi}{8}$ and $\frac{g_{1}}{g_{2}}=\frac{3}{\sqrt{2}}$ while the solid line corresponds to a 2-particle excitation where each of the particles occupies the state $\ket{\phi_{B}}=\frac{1}{\sqrt{2}}(\ket{\phi_{1}}+\ket{\phi_{2}}).$}
\label{F:2particles}
\end{figure}

From the above equations, one finds that the maximum value of entanglement in $A_{1}A_{2}$ is achieved when $|\sin(4\phi)|=1$.
 It is interesting to compare this situation with the case when both particles occupy the same state (see (\ref{Nkk}) and (\ref{N23})); if $g_{1} \neq g_{2}$, one can increase the entanglement transferred to the qubits preparing
   B in a state of the form $(\ref{su(2)})$  (see Fig.{\ref{F:2particles}}).

\section{Conclusions}
 We have studied the entanglement induced in a two qubit system as a result of its interaction with a bosonic system. The operators coupling each of the  qubits to the bosonic system were assumed to commute. As discussed throughout this paper, these interactions appear in the situation when one couples each qubit to a  different mode.  More precisely, we considered operators of the form
  \beq
  \label{classy}
  \mF_{i}=p_{i}({a_{i}}^{\dagger}a_{i}){a_{i}}^{n}+q_{i}(a_{i}^{\dagger}a_{i}){a_{i}^{\dagger}}^{m}, \quad i=(1,2)\quad \textrm{with} \quad [a_{1},{a_{2}}^{\dagger}]=0.
 \eeq

In this case, the mechanism entangling the qubits is analogous to the mechanism responsible for the entanglement transfer from two qubit systems to two qubit systems (see Fig.\ref{F:qubit-qubit}). From  section (\ref{S:N}), we know that  a 1-particle state being a superposition of  the modes $\ket{\phi_{1}}$, $\ket{\phi_{2}}$  takes the form of the entangled state
\beq
\ket{\Phi_{B}}=a^{\dagger}(\phi_{B})\ket{0}=u_{1}\ket{1,0}+u_{2}\ket{0,1}
\eeq
when written in occupation number representation. However, the form of state $\ket{\Phi_{B}}$ depends on the interaction between the qubits and system B. If one of the qubits interacts with mode $\ket{\phi_{B}}$ while the other qubit interacts with  mode $\ket{\phi_{B}'}$ (orthogonal to $\ket{\phi_{B}}$),  then the state $\ket{\Phi_{B}}=a^{\dagger}(\phi_{B})\ket{0}$ may be written as
\beq
\ket{\Phi_{B}}=\ket{1}\ket{0}.
\eeq
Now, $\ket{\Phi_{B}}$ has the form of a separable state. It is for this reason that we avoided talking about the entanglement between the modes. Instead, we computed the entanglement induced in the two qubit system as a result of the interaction with multiparticle systems. For all the N-particles states considered, we found an interaction inducing entanglement in the two qubit system. This situation changes dramatically if the bosonic system is in
 the coherent state  $\ket{z}=e^{z a(\phi_{B})^{\dagger}-z^{*}a(\phi_{B})}\sim e^{z a^{\dagger}(\phi_{B})}\ket{0} $. In fact, this state behaves like a separable state for operators of the form $(\ref{classy})$. In section ({\ref{S:Expansion}}),  we studied the series expansion of the negativity $\mathcal{N}({\rho^{A}})$. We computed the first nonvanishing contribution to $\mathcal{N}(\rho^{A})$ in the case where the operators acting on B were different from those in ($\ref{classy}$). We found that when system B is in the particle vacuum state $\ket{0}$, the qubits may become entangled if the interaction Hamiltonian contains  operators of the form  $\mF_{i}=g_{i}(a(\phi_{i})+\beta_{i}a^{\dagger}(\psi_{i}))$. This type of interactions could be used to extract entanglement from a coherent state $\ket{z}$ (entanglement extraction from coherent  states has been discussed in \cite{Vedral}). We leave the this problem for future work.

 \begin{center}
 {\bf ACKNOWLEDGEMENTS}
 \end{center}
The author is grateful to Professors Thomas Curtright and Luca Mezincescu for helpful comments. He would also like to thank Lukasz Cywinski and Dan Pruteanu for their interest in this work.

\end{document}